\documentclass[a4paper,11pt]{article} 

\pdfoutput=1

\usepackage{physics}
\usepackage{hyperref}
\usepackage{enumitem}
\usepackage{xspace}
\usepackage{slashed}
\usepackage{braket}
\usepackage{leftindex}
\usepackage{graphicx}  
\usepackage{bm}  
\usepackage{booktabs}
\usepackage[normalem]{ulem}
\usepackage{graphics}
\usepackage{color}
\usepackage{colortbl}
\usepackage[svgnames,dvipsnames,table,x11names]{xcolor}
\usepackage{comment}
\usepackage{enumitem}
\usepackage{mathrsfs}
\usepackage{adjustbox}
\usepackage{hhline}
\usepackage{lmodern}

\usepackage{jheppub} 

\hypersetup{
    pdfencoding=unicode,
	colorlinks=true,
	urlcolor=RoyalBlue,
	linkcolor=Maroon,
	citecolor=ForestGreen,
    linktoc=page,
    linktocpage=true,
}

\newcommand{\agl}[2]{\langle #1 #2 \rangle}
\newcommand{\sqr}[2]{\lbrack #1 #2 \rbrack}

\def\coef#1#2#3{ {\vphantom{A^2}}^{#1}\!{{C}}^{#3}_{#2} }
\def\op#1#2#3{ {\vphantom{A^2}}^{#1}\!{{Q}}^{#3}_{#2} }

\DeclareDocumentCommand\derivative{ s o m g d() }
{ 
    \IfBooleanTF{#1}
    {\let\fractype\flatfrac}
    {\let\fractype\frac}
    \IfNoValueTF{#4}
    {
        \IfNoValueTF{#5}
        {\fractype{\diffd \IfNoValueTF{#2}{}{^{#2}}}{\diffd #3\IfNoValueTF{#2}{}{^{#2}}}}
        {\fractype{\diffd \IfNoValueTF{#2}{}{^{#2}}}{\diffd #3\IfNoValueTF{#2}{}{^{#2}}} \argopen(#5\argclose)}
    }
    {\fractype{\diffd \IfNoValueTF{#2}{}{^{#2}} #3}{\diffd #4\IfNoValueTF{#2}{}{^{#2}}}\IfValueT{#5}{(#5)}}
}

\makeatletter\g@addto@macro\bfseries{\boldmath}\makeatother

\title{Positivity and partial wave unitarity bounds on ALP theories via amplitude methods}

\author[a,b]{Luigi~C.~Bresciani,}
\author[c]{Gabriele~Levati}
\author[a,b]{and Paride~Paradisi}
\affiliation[a]{Dipartimento di Fisica e Astronomia ``G.~Galilei'', Universit\`a degli Studi di Padova,\\Via F.~Marzolo 8, I-35131 Padova, Italy}
\affiliation[b]{Istituto Nazionale di Fisica Nucleare, Sezione di Padova, Via F.~Marzolo 8, I-35131 Padova, Italy}
\affiliation[c]{Albert Einstein Center for Fundamental Physics, Institute for Theoretical Physics, University of Bern, Sidlerstrasse 5, CH-3012 Bern, Switzerland}

\emailAdd{luigicarlo.bresciani@phd.unipd.it}
\emailAdd{gabriele.levati@unibe.ch}
\emailAdd{paride.paradisi@unipd.it}

\abstract{We derive the complete set of partial wave unitarity bounds on the most general Axion-Like Particle (ALP) effective interactions up to dimension 8 in the limit of large center-of-mass energy. Exploiting a recently developed formalism based on spinor-helicity techniques, we discuss the unitarity bounds for $N \to M$ (with $N, M \geq 2$) scattering amplitudes that can be relevant for ALP searches at colliders or in a variety of rare processes. Moreover, we compute positivity bounds on ALP interactions, emphasizing their complementarity with partial wave unitarity bounds. As a byproduct, we show that our results can be used to infer new positivity constraints in the Standard Model Effective Field Theory.}

\allowdisplaybreaks

\begin{document}
\maketitle
\flushbottom
\pagestyle{myplain}

\section{Introduction}

Axion-like particles (ALPs) are light spin-0 bosons emerging in several extensions of the Standard Model (SM) after the spontaneous breaking of an underlying global symmetry at energies much larger than the electroweak scale~\cite{Jaeckel:2010ni,Marsh:2015xka,Irastorza:2018dyq,DiLuzio:2020wdo}.
Therefore, their lightness can be naturally explained by their pseudo-Nambu-Goldstone boson (pNGB) nature.
ALPs are a generalization of the QCD axion, as their mass and symmetry breaking scale are independent parameters to be probed by experiment. 
Remarkably, ALPs can answer several open questions in particle physics, such as the strong CP problem
\cite{Peccei:1977hh,Peccei:1977ur,Weinberg:1977ma,Wilczek:1977pj}, the flavor puzzle \cite{Davidson:1981zd,Wilczek:1982rv,Berezhiani:1989fp,Calibbi:2016hwq,Greljo:2024evt}, and the electroweak scale naturalness~\cite{Graham:2015cka}. Furthermore, ALPs can be natural dark matter candidates~\cite{Abbott:1982af,Preskill:1982cy,Dine:1982ah,Davis:1986xc}.
ALPs with masses below the MeV scale can leave their imprints in several cosmological and astrophysical experimental searches~\cite{ADMX:2003rdr,Barbieri:2016vwg,Caldwell:2016dcw,Zioutas:1998cc,Irastorza:2011gs,CAST:2017uph,Armengaud:2014gea,VanBibber:1987rq,Bahre:2013ywa,OSQAR:2015qdv,Arvanitaki:2014dfa,Alekhin:2015byh,Dobrich:2015jyk}. Instead, larger ALP masses can be explored at colliders and through a variety of rare processes~\cite{Bauer:2017ris,Bauer:2021mvw,Gavela:2019wzg,Cornella:2019uxs,DiLuzio:2023ndz,Marciano:2016yhf,DiLuzio:2020oah,DiLuzio:2023cuk}.
ALP interactions with SM fields are typically treated model-independently by means of dimension-5 effective operators~\cite{Georgi:1986df}. 
This approach has the advantage of capturing general features of large classes of ultraviolet models without sticking to any particular scenario.
Quite recently, a systematic classification of higher-dimensional ALP 
operators --- up to dimension 8 --- appeared in Ref.~\cite{Bertuzzo:2023slg}
exploiting on-shell techniques~\cite{DeAngelis:2022qco,Li:2022tec}.

The pNGB nature of ALPs forbids non-derivative interactions with SM fields. As a result, ALP interactions show an inherent growth with energy, making them particularly prone to experience strong unitarity constraints.
In particular, partial wave unitarity bounds --- stemming from general properties 
of the scattering $S$-matrix --- provide a theoretically robust way to infer upper limits on the strength of such interactions, thus complementing experimental bounds. 
Historically, the emergence of unitarity violation for the $WW$ scattering 
process led to an upper bound on the Higgs boson mass below the TeV scale~\cite{Lee:1977yc,Lee:1977eg}, motivating the construction of the LHC. 
More recently, unitarity bounds have also been extensively discussed in the context of Effective Field Theories (EFTs)~\cite{Gounaris:1994cm,Corbett:2014ora,Corbett:2017qgl,DiLuzio:2017chi,DiLuzio:2016sur,Mahmud:2024iyn,Allwicher:2021jkr,Cohen:2021ucp,Almeida:2020ylr,Cohen:2021gdw,Abu-Ajamieh:2020yqi,Brivio:2021fog}.
These bounds are inherently perturbative in nature and signal the scale
at which new dynamics is expected
to emerge, either via additional degrees of freedom or the onset of a
strongly coupled phase.

The standard approach to calculate them exploits $2 \to 2$ scattering processes 
of particles with helicities $h_i$ and proceeds through the following steps: 
i) expansion of the helicity amplitudes into partial waves $a_{h_i}^J$ with total angular momentum $J$ using the Wigner rotation matrix~\cite{Jacob:1959at} and ii) diagonalization of the partial wave scattering matrix.
As a result, the tightest limit arises from the partial wave unitarity bound $|a_{h_i}^J|_{\rm max} \lesssim 1$ on the largest eigenvalue.
However, the above method is not suited to treat either $2 \to N$ (with $N > 2$) amplitudes, which are of great relevance for future high-energy colliders, 
or spin-2 or higher-spin theories, relevant for EFTs of gravity, given the 
difficulty in calculating  amplitudes via Feynman rules.
By contrast, on-shell methods --- which have been shown to be very efficient in capturing EFT ultraviolet effects via renormalization group equations~\cite{Caron-Huot:2016cwu,EliasMiro:2020tdv,Baratella:2020lzz,Jiang:2020mhe,Bern:2020ikv,Baratella:2020dvw,AccettulliHuber:2021uoa,EliasMiro:2021jgu,Baratella:2022nog,Machado:2022ozb,Bresciani:2023jsu,Bresciani:2024shu,Aebischer:2025zxg} --- turned out to be the ideal tool also to account for unitarity bounds of EFTs at high energies~\cite{Bresciani:2025toe}.
In particular, building on a vectorial formalism~\cite{Jiang:2020rwz} based on spinor-helicity techniques~\cite{Dixon:2013uaa}, in Ref.~\cite{Bresciani:2025toe} we provided the general angular momentum basis for $2 \to 3$ amplitudes thanks to the amplitude-operator correspondence~\cite{Shadmi:2018xan,Durieux:2019eor,Dong:2021vxo,Li:2022tec}. 
Moreover, we discussed unitarity bounds for some dimension-6 and -8
operators in the Standard Model Effective Field Theory (SMEFT) as well as for spin-2 or higher-spin theories. 

The aim of this work is to extensively apply the method of Ref.~\cite{Bresciani:2025toe} to ALP EFTs up to dimension 8,
improving and generalizing previous results in the literature. 
In particular, we revisit the bounds of Ref.~\cite{Brivio:2021fog} 
relative to dimension-5 operators and include those for weak-violating ALP interactions, which exhibit energy enhancements in various processes such as charged meson decays and $W$ boson decays~\cite{Altmannshofer:2022ckw}. Moreover, we address the unitarity bounds for higher-dimensional operators up to dimension 8 recently classified in~\cite{Song:2023lxf,Grojean:2023tsd,Bertuzzo:2023slg}.

Finally, we evaluate the complete set of positivity bounds
for ALP EFTs, discussing their complementarity and interplay with partial wave unitarity constraints~\cite{Bresciani:2025toe}.
Positivity bounds provide a powerful handle to infer the parameter space of consistent EFTs as they follow from requiring the EFT to be the low-energy limit of a unitary, local, and causal quantum field theory~\cite{Adams:2006sv}. 
They have been applied across a broad range of frameworks, including the SMEFT \cite{Bellazzini:2018paj,Zhang:2020jyn,Bonnefoy:2025uzf,Bi:2019phv,Remmen:2019cyz,Remmen:2020vts,Fuks:2020ujk,Yamashita:2020gtt,Gu:2020ldn,Bonnefoy:2020yee,Henriksson:2021ymi,Zhang:2021eeo,Chala:2021wpj,Chala:2023jyx,Chala:2023xjy,Li:2022rag,Ghosh:2022qqq,Li:2022aby,Gu:2023emi,Chen:2023bhu,Ye:2024rzr,Ye:2025zhs,Davighi:2023acq}, Higgs EFT \cite{Remmen:2024hry,Chakraborty:2024ciu}, and gravitational theories \cite{Cheung:2016yqr,Bellazzini:2017fep,deRham:2017xox,Alberte:2020jsk,Hong:2023zgm,Alviani:2024sxx,Tokuda:2020mlf,Alberte:2020bdz,Chiang:2022jep,deRham:2022gfe,Caron-Huot:2021rmr,Alberte:2021dnj,Caron-Huot:2022ugt}.

In the following, we outline the structure of the paper. 
In Section~\ref{review_method} we review the method for computing partial wave unitarity bounds in the on-shell approach. 
In Section~\ref{ALP_EFT} we present the ALP EFT, including effective operators up to dimension 8. 
In Section~\ref{unitarity_bounds} we systematically present the partial wave unitarity bounds in our scenario, emphasizing the important role of a coupled-channel analysis.
In Section~\ref{sec:positivity} we discuss positivity bounds for dimension-8 operators and their complementarity with partial wave unitarity bounds. Section~\ref{pheno} is dedicated to some phenomenological applications of our results. 
Our conclusions are discussed in Section~\ref{conclusions}. 
In Appendix~\ref{sec:conventions} we list the conventions adopted throughout the paper,
in Appendix~\ref{Appendix_A} we list the relevant amplitudes to obtain the most stringent partial wave unitarity bounds reported in the main text,
while in Appendix~\ref{Appendix_B} we
test the obtained positivity bounds against a specific UV extension.

\section{Partial wave unitarity bounds in the on-shell approach}
\label{review_method}

In this section, we review the method for computing partial wave unitarity bounds in the on-shell formalism~\cite{Bresciani:2025toe}. 
The partial wave analysis enables one to project a generic amplitude $\ket{\mathcal A_{i\to f}}$ onto a kinematic basis $\ket{\mathcal B^J_{i\to f}}$ with definite angular momentum $J$ as follows
\begin{equation}
    \ket{\mathcal A_{i\to f}} = \sum_J a^J_{i\to f} \ket{\mathcal B^J_{i\to f}}\,,
    \label{eq:pwd}
\end{equation}
where $a^J_{i\to f}$ are the partial wave coefficients.
The building blocks of this decomposition are the Poincaré Clebsch-Gordan coefficients $\mathcal C_{\mathcal I \to *}^{J,h}$ defined as~\cite{Jiang:2020rwz}
\begin{equation}
    \langle{P,J,h|\mathcal I}\rangle = \mathcal C_{\mathcal I \to *}^{J,h}\,
    \delta^{(4)}\left(P-\sum_{i\in\mathcal I}p_i\right)\,,
\end{equation}
that is, the overlap between the multiparticle state $\ket{\mathcal I}$ and the Poincaré irreducible multiparticle state $\ket{P,J,h}$, where $h$ refers to its helicity.
The coefficients $\mathcal C_{\mathcal I \to *}^{J,h}$ can be interpreted as elements of the complex vector space $V_{\mathcal I \to *}$, i.e., $\ket{\mathcal C_{\mathcal I \to *}^{J,h}}\in V_{\mathcal I \to *}$, and, correspondingly, the angular momentum basis elements read
\begin{equation}
    \ket{\mathcal B^{J}_{i\to f}} = \sum_h \ket{\mathcal C_{i \to *}^{J,h}} \otimes \ket{\mathcal C_{* \to f}^{J,h}} \in V_{i\to f}\,,
\end{equation}
where
$\ket{\mathcal C^{J,h}_{* \to \mathcal I}}=\ket{(\mathcal C^{J,h}_{\mathcal I \to *})^*} \in V_{\ast \to \mathcal I}$ and $V_{i\to f} = V_{i \to *}\otimes V_{* \to f}$.
The partial wave coefficients $a_{i\to f}^J$ can then conveniently be determined by extending the inner product in $V_{\mathcal I \to *}$~\cite{Shu:2021qlr} to the vector space $V_{i\to f}$ as follows
\begin{equation}
    a_{i\to f}^J = \frac{1}{2J+1}\langle \mathcal B^J_{i\to f}|\mathcal A_{i\to f}\rangle 
    =
    \frac{1}{2J+1}\int \text{d}\Phi_i\, \text{d}\Phi_f\, \left(\mathcal B^J_{i\to f}\right)^* \mathcal A_{i\to f}\,,
    \label{eq:partial_wave}
\end{equation}
where $\text{d}\Phi_{\mathcal I} =  \frac{1}{S_{\mathcal{I}}} \prod_{k \in \mathcal{I}} \frac{\dd^3 p_k}{(2\pi)^3 2E_k}$ is the Lorentz-invariant phase space measure related to $\ket{\mathcal I}$
and $S_{\mathcal I}$ is the symmetry factor accounting for identical particles in $\ket{\mathcal{I}}$.\footnote{Explicit parameterizations for the $n$-body phase space in terms of spinor-helicity variables can be found in \cite{Zwiebel:2011bx,Mastrolia:2009dr,Cox:2018wce,Larkoski:2020thc,EliasMiro:2020tdv}.}
Notice that the adopted basis normalization
$\langle \mathcal B^J_{i\to f}|\mathcal B^{J'}_{i\to f}\rangle = (2J+1)\delta^{JJ'}$
has been chosen to obtain the partial wave unitarity bounds in the standard form.
Indeed, this choice implies
\begin{equation}
    \int \text d \Phi_X\,\ket{\mathcal B^{J}_{i\to X}}\otimes \ket{\mathcal B_{X\to f}^{J'}} = \ket{\mathcal B_{i\to f}^J} \delta^{JJ'}\,,
\end{equation}
which, when applied to the generalized optical theorem
\begin{equation}
    \ket{\mathcal A_{i\to f}} - \ket{\mathcal A_{f\to i}^*} = i \sum_X \int \text d \Phi_X\, \ket{\mathcal A_{i\to X}}\otimes\ket{\mathcal A_{f\to X}^* }\,,
\end{equation}
leads to  
\begin{equation}
    a_{i\to f}^J - \left(a_{f\to i}^J\right)^* = i \sum_X a_{i\to X}^{J} \left(a_{f\to X}^{J}\right)^*
\end{equation}
and therefore to the partial wave unitarity bounds
\begin{equation}
    |\! \Re a_{i\to i}^J |  \le 1 \,,
    \qquad
    0 \le \Im a_{i\to i}^J \le 2\,,
    \qquad 
    | a^J_{i\to f} | \le 1 \,,
\end{equation}
where $i\neq f$.
The angular momentum basis elements $\ket{\mathcal B^{J}_{i\to f}}$, required by the partial wave decomposition, can be determined without passing through the Poincaré Clebsch-Gordan coefficients $\ket{\mathcal C_{\mathcal I \to *}^{J,h}}$, via the following steps:
\begin{enumerate}
    \item Determine a set of kinematic monomials in terms of spinor-helicity variables that are consistent with particle helicities, span $V_{i\to f}$, and are mutually independent after accounting for momentum conservation and Schouten identities.\footnote{Public Mathematica packages allowing one to achieve this goal can be found in~\cite{DeAngelis:2022qco,Li:2022tec}.}
    \item Apply the Pauli-Lubanski operator squared~\cite{Witten:2003nn}
    \begin{equation}
    \mathbb W^2_{\mathcal I} = \frac{1}{8}\mathbb P_{\mathcal I}^2 \left(\epsilon^{\alpha\gamma}\epsilon^{\beta\delta}\mathbb M_{\mathcal I,\alpha\beta}\mathbb M_{\mathcal I,\gamma\delta}+\epsilon^{\dot\alpha\dot\gamma}\epsilon^{\dot\beta\dot\delta}\widetilde{\mathbb M}_{\mathcal I,\dot\alpha\dot\beta}\widetilde{\mathbb M}_{\mathcal I,\dot\gamma\dot\delta}\right)
    +\frac{1}{4}\mathbb P_{\mathcal I}^{\alpha\dot\alpha}\mathbb P_{\mathcal I}^{\beta\dot\beta}\mathbb M_{\mathcal I,\alpha\beta}\widetilde{\mathbb M}_{\mathcal I,\dot\alpha\dot\beta}\,,
\end{equation}
to these monomials and construct the corresponding matrix. In the previous expression, $\mathcal I=i,f$ and
\begin{align}
    \mathbb P_{\mathcal I}^{\alpha\dot\alpha} &= \sum_{i\in \mathcal I}\lambda_i^\alpha \widetilde \lambda_i^{\dot\alpha}\,,\\
    \mathbb M_{\mathcal I}^{\alpha\beta} &= \sum_{i\in \mathcal I}\left(\lambda_{i}^\alpha\frac{\partial}{\partial \lambda_{i,\beta}}+\lambda_{i}^\beta \frac{\partial}{\partial \lambda_{i,\alpha}}\right)\,,\\
    \widetilde{\mathbb M}_{\mathcal I}^{\dot\alpha\dot\beta} &= \sum_{i\in \mathcal I}\left(\widetilde\lambda_{i}^{\dot\alpha}\frac{\partial}{\partial \widetilde\lambda_{i,\dot\beta}}+\widetilde\lambda_{i}^{\dot\beta}\frac{\partial}{\partial \widetilde\lambda_{i,\dot\alpha}}\right)\,.
\end{align}
\item Find the eigenvectors of the above matrix and the associated angular momentum values $J_{\mathcal I}$ from the Casimir eigenvalues $-P^2_{\mathcal I}J_{\mathcal I}(J_{\mathcal I}+1)$.
\item Normalize the elements of the determined orthogonal basis so that their norm is $\sqrt{2 J_{\mathcal I}+1}$.
\end{enumerate}

As a consistency check, for $2\to 2$ scattering, the basis $\ket{\mathcal B^{J}_{i\to f}}$ turns out to be proportional to the Wigner $d$-matrix~\cite{Jiang:2020rwz}. 
Moreover, the partial wave coefficients for $N\to M$ scattering (with $N,M \geq 2$) reproduce the elements of the reduced scattering matrix~\cite{Chang:2019vez,Falkowski:2019tft,Cohen:2021ucp,Mahmud:2025wye}
for the $s$-wave contributions~\cite{Bresciani:2025toe}.

\section{The ALP effective field theory}
\label{ALP_EFT}

ALP interactions with SM fermions and gauge bosons 
are generated starting from dimension-5 effective operators \cite{Georgi:1986df}. 
The ALP EFT Lagrangian takes the form
\begin{equation}
    \mathcal L_{\phi} = \frac{1}{2}\partial_\mu \phi \, \partial^\mu \phi + \sum_{d,k}\coef{d}{k}{}\op{d}{k}{}
\end{equation}
with $d = [\op{d}{k}{}] = 4- [\coef{d}{k}{}]\in\{5,6,7,8\}$.
The complete set of operators invariant under a shift symmetry has been systematically classified in \cite{Song:2023lxf,Grojean:2023tsd,Bertuzzo:2023slg} up to mass dimension 8 and is reported in Tables \ref{tab:dim5-6}, \ref{tab:dim7}, and \ref{tab:dim8}.

\begin{table}[h!tb]
\centering
\setlength{\tabcolsep}{1pt}
\renewcommand{\arraystretch}{1.5}
\begin{tabular}{ c|c||c|c||c|c }
\toprule
\multicolumn{2}{c||}{\cellcolor{gray!20}\hyperref[par:phipsi2H]{\boldmath$\phi\psi^2 H+\text{\bf h.c.}$}} & \multicolumn{2}{c||}{\cellcolor{gray!20}\hyperref[par:phiX2]{\boldmath$\phi X^2$}} & \multicolumn{2}{c}{\cellcolor{gray!20}\hyperref[par:phi2H2D2]{\boldmath$\phi^2 H^2 D^2$}} \\
\hline
$\op{5}{\phi \ell e H}{pr}$ & $\phi\, \overline \ell_p  e_r H$ & $\op{5}{\phi B^2}{}$ & $\phi\, B_{\mu\nu}\widetilde B^{\mu\nu}$&$\op{6}{\phi^2 H^2}{}$& $(\partial_\mu \phi\, \partial^\mu \phi)(H^\dagger H)$ 
\\ 
$\op{5}{\phi q u H}{pr}$ & $\phi\, \overline q_p  u_r \widetilde H$ & $\op{5}{\phi W^2}{}$ & $\phi\, W^I_{\mu\nu}\widetilde W^{I \mu\nu}$ &\multicolumn{2}{c}{}
\\
$\op{5}{\phi q d H}{pr}$ & $\phi\, \overline q_p  d_r H$ & $\op{5}{\phi G^2}{}$ & $\phi\, G^A_{\mu\nu}\widetilde G^{A\mu\nu}$ &\multicolumn{2}{c}{}
\\
\bottomrule
\end{tabular}
\caption{Dimension-5 and -6 ALP operators.
Operators with $+ \text{h.c.}$ have Hermitian conjugates.
The indices $p,r$ denote weak eigenstates. The labels of the operator classes contain a cross-reference to the corresponding unitarity bounds.\label{tab:dim5-6}}
\end{table}

\begin{table}[h!tb]
\centering
\setlength{\tabcolsep}{1pt}
\renewcommand{\arraystretch}{1.5}
\begin{tabular}{ c|c||c|c }
\toprule
\multicolumn{2}{c||}{\cellcolor{gray!20}\hyperref[par:phipsi2XD]{\boldmath$\phi \psi^2 X D$}} & \multicolumn{2}{c}{\cellcolor{gray!20}\hyperref[par:phiXH2D2]{\boldmath$\phi X H^2 D^2$}} \\
\hline
$\op{7}{\phi \ell^2 B}{(1)pr}$ & $(\partial^\mu \phi)(\overline \ell_p \gamma^\nu \ell_r)B_{\mu\nu}$ & $\op{7}{\phi B H^2}{(1)}$ & $i(\partial_\mu \phi)(H^\dagger \overset{\leftrightarrow}{D}{}_\nu H)B^{\mu\nu}$
\\
$\op{7}{\phi \ell^2 B}{(2)pr}$ & $(\partial^\mu \phi)(\overline \ell_p \gamma^\nu \ell_r)\widetilde B_{\mu\nu}$  & $\op{7}{\phi B H^2}{(2)}$ & $i(\partial_\mu \phi)(H^\dagger \overset{\leftrightarrow}{D}{}_\nu H) \widetilde B^{\mu\nu}$
\\
$\op{7}{\phi e^2 B}{(1)pr}$ & $(\partial^\mu \phi)(\overline e_p \gamma^\nu e_r)B_{\mu\nu}$ & $\op{7}{\phi W H^2}{(1)}$ & $i(\partial_\mu \phi)(H^\dagger \overset{\leftrightarrow}{D}{}_\nu^I H)W^{I\mu\nu}$
\\
$\op{7}{\phi e^2 B}{(2)pr}$ & $(\partial^\mu \phi)(\overline e_p \gamma^\nu e_r)\widetilde B_{\mu\nu}$  & $\op{7}{\phi W H^2}{(2)}$ & $i(\partial_\mu \phi)(H^\dagger \overset{\leftrightarrow}{D}{}^I_\nu H) \widetilde W^{I\mu\nu}$
\\ \hhline{~|~||--}
$\op{7}{\phi q^2 B}{(1)pr}$ & $(\partial^\mu \phi)(\overline q_p \gamma^\nu q_r)B_{\mu\nu}$ & \multicolumn{2}{c}{\cellcolor{gray!20}\hyperref[par:phipsi2H2D]{\boldmath$\phi \psi^2 H^2 D$}} 
\\ \hhline{~|~||--}
$\op{7}{\phi q^2 B}{(2)pr}$ & $(\partial^\mu \phi)(\overline q_p \gamma^\nu q_r)\widetilde B_{\mu\nu}$   & $\op{7}{\phi \ell^2 H^2}{(1)pr}$ & $(\partial_\mu \phi)(\overline \ell_p \gamma^\mu \ell_r)(H^\dagger H)$ 
\\
$\op{7}{\phi u^2 B}{(1)pr}$ & $(\partial^\mu \phi)(\overline u_p \gamma^\nu u_r)B_{\mu\nu}$  & $\op{7}{\phi \ell^2 H^2}{(2)pr}$ & $(\partial_\mu \phi)(\overline \ell_p \gamma^\mu \sigma^I \ell_r)(H^\dagger \sigma^I H)$
\\
$\op{7}{\phi u^2 B}{(2)pr}$ & $(\partial^\mu \phi)(\overline u_p \gamma^\nu u_r)\widetilde B_{\mu\nu}$  & $\op{7}{\phi q^2 H^2}{(1)pr}$ & $(\partial_\mu \phi)(\overline q_p \gamma^\mu q_r)(H^\dagger H)$
\\
$\op{7}{\phi d^2 B}{(1)pr}$ & $(\partial^\mu \phi)(\overline d_p \gamma^\nu d_r)B_{\mu\nu}$ & $\op{7}{\phi q^2 H^2}{(2)pr}$ & $(\partial_\mu \phi)(\overline q_p \gamma^\mu \sigma^I q_r)(H^\dagger \sigma^I H)$
\\
$\op{7}{\phi d^2 B}{(2)pr}$ & $(\partial^\mu \phi)(\overline d_p \gamma^\nu d_r)\widetilde B_{\mu\nu}$  & $\op{7}{\phi e^2 H^2}{pr}$ & $(\partial_\mu \phi)(\overline e_p \gamma^\mu e_r)(H^\dagger H)$
\\
$\op{7}{\phi \ell^2 W}{(1)pr}$ & $(\partial^\mu \phi)(\overline \ell_p \gamma^\nu \sigma^I \ell_r)W^I_{\mu\nu}$ & $\op{7}{\phi u^2 H^2}{pr}$ & $(\partial_\mu \phi)(\overline u_p \gamma^\mu u_r)(H^\dagger H)$
\\
$\op{7}{\phi \ell^2 W}{(2)pr}$ & $(\partial^\mu \phi)(\overline \ell_p \gamma^\nu \sigma^I \ell_r)\widetilde W^I_{\mu\nu}$ & $\op{7}{\phi d^2 H^2}{pr}$ & $(\partial_\mu \phi)(\overline d_p \gamma^\mu d_r)(H^\dagger H)$
\\ \hhline{~|~||--}
$\op{7}{\phi q^2 W}{(1)pr}$ & $(\partial^\mu \phi)(\overline q_p \gamma^\nu \sigma^I q_r)W^I_{\mu\nu}$ & \multicolumn{2}{c}{\cellcolor{gray!20}\hyperref[par:phipsi2HD2]{\boldmath$\phi \psi^2 H D^2 + \text{\bf h.c.}$}} 
\\ \hhline{~|~||--}
$\op{7}{\phi q^2 W}{(2)pr}$ & $(\partial^\mu \phi)(\overline q_p \gamma^\nu \sigma^I q_r) \widetilde W^I_{\mu\nu}$ & $\op{7}{\phi \ell e H}{(1)pr}$ & $(\partial_\mu \phi)(\overline \ell_p D^\mu e_r H)$
\\
$\op{7}{\phi q^2 G}{(1)pr}$ & $(\partial^\mu \phi)(\overline q_p \gamma^\nu \lambda^A q_r)G^A_{\mu\nu}$ & $\op{7}{\phi \ell e H}{(2)pr}$ & $(\partial_\mu \phi)(D^\mu \overline \ell_p e_r H)$ 
\\
$\op{7}{\phi q^2 G}{(2)pr}$ & $(\partial^\mu \phi)(\overline q_p \gamma^\nu \lambda^A q_r) \widetilde G^A_{\mu\nu}$ & $\op{7}{\phi qu H}{(1)pr}$ & $(\partial_\mu \phi)(\overline q_p D^\mu u_r \widetilde H)$ 
\\
$\op{7}{\phi u^2 G}{(1)pr}$ & $(\partial^\mu \phi)(\overline u_p \gamma^\nu \lambda^A u_r)G^A_{\mu\nu}$ & $\op{7}{\phi qu H}{(2)pr}$ & $(\partial_\mu \phi)(D^\mu \overline q_p u_r \widetilde H)$ 
\\
$\op{7}{\phi u^2 G}{(2)pr}$ & $(\partial^\mu \phi)(\overline u_p \gamma^\nu \lambda^A u_r)\widetilde G^A_{\mu\nu}$ & $\op{7}{\phi qd H}{(1)pr}$ & $(\partial_\mu \phi)(\overline q_p D^\mu d_r  H)$
\\
$\op{7}{\phi d^2 G}{(1)pr}$ & $(\partial^\mu \phi)(\overline d_p \gamma^\nu \lambda^A d_r)G^A_{\mu\nu}$ & $\op{7}{\phi qd H}{(2)pr}$ & $(\partial_\mu \phi)(D^\mu \overline q_p d_r H)$ 
\\ \hhline{~|~||--}
$\op{7}{\phi d^2 G}{(2)pr}$ & $(\partial^\mu \phi)(\overline d_p \gamma^\nu \lambda^A d_r)\widetilde G^A_{\mu\nu}$ & \multicolumn{2}{c}{\cellcolor{gray!20}\hyperref[par:phiH4D2]{\boldmath$\phi H^4 D^2$}} 
\\  \hhline{~|~||--}
\multicolumn{2}{c||}{} & $\op{7}{\phi H^4}{}$ & $i(\partial^\mu \phi) (H^\dagger \overset{\leftrightarrow}{D}{}_\mu H)(H^\dagger H)$
\\ 
\bottomrule
\end{tabular}
\caption{Dimension-7 ALP operators. Operators with $+ \text{h.c.}$ have Hermitian conjugates. The indices $p,r$ denote weak eigenstates.
$\sigma^I$ are the Pauli matrices and $\lambda^A$ are the Gell-Mann matrices. The labels of the operator classes contain a cross-reference to the corresponding unitarity bounds.\label{tab:dim7}}
\end{table}

\begin{table}[h!tb]
\centering
\setlength{\tabcolsep}{1pt}
\renewcommand{\arraystretch}{1.5}
\begin{tabular}{ c|c||c|c }
\toprule
\multicolumn{2}{c||}{\cellcolor{gray!20}\hyperref[par:phi2X2D2]{\boldmath$\phi^2 X^2 D^2$}} & \multicolumn{2}{c}{\cellcolor{gray!20}\hyperref[par:phi2psi2D3]{\boldmath$\phi^2 \psi^2 D^3$}} 
\\
\hline
$\op{8}{\phi^2 B^2}{(1)}$ & $(\partial^\mu \phi\, \partial_\nu \phi)B_{\mu\rho}B^{\nu \rho}$ & $\op{8}{\phi^2 \ell^2}{pr}$ & $i(\partial_\mu \phi\, \partial_\nu \phi)(\overline \ell_p \gamma^\mu D^\nu \ell_r)$
\\
$\op{8}{\phi^2 B^2}{(2)}$ & $(\partial_\mu \phi \, \partial^\mu \phi)B_{\nu\rho}B^{\nu\rho}$ & $\op{8}{\phi^2 e^2}{pr}$ & $i(\partial_\mu \phi\, \partial_\nu \phi)(\overline e_p \gamma^\mu D^\nu e_r)$ 
\\
$\op{8}{\phi^2 B^2}{(3)}$ & $(\partial_\mu \phi \, \partial^\mu \phi)B_{\nu\rho}\widetilde B^{\nu\rho}$ & $\op{8}{\phi^2 q^2}{pr}$ & $i(\partial_\mu \phi\, \partial_\nu \phi)(\overline q_p \gamma^\mu D^\nu q_r)$
\\ 
$\op{8}{\phi^2 W^2}{(1)}$ & $(\partial^\mu \phi\, \partial_\nu \phi)W_{\mu\rho}^I W^{I \nu \rho}$ & $\op{8}{\phi^2 u^2}{pr}$ & $i(\partial_\mu \phi\, \partial_\nu \phi)(\overline u_p \gamma^\mu D^\nu u_r)$
\\ 
$\op{8}{\phi^2 W^2}{(2)}$ & $(\partial_\mu \phi \, \partial^\mu \phi)W_{\nu\rho}^I W^{I\nu\rho}$ & $\op{8}{\phi^2 d^2}{pr}$ & $i(\partial_\mu \phi\, \partial_\nu \phi)(\overline d_p \gamma^\mu D^\nu d_r)$
\\ \hhline{~|~||--}
$\op{8}{\phi^2 W^2}{(3)}$ & $(\partial_\mu \phi \, \partial^\mu \phi)W_{\nu\rho}^I \widetilde W^{I\nu\rho}$ & \multicolumn{2}{c}{\cellcolor{gray!20}\hyperref[par:phi2psi2HD2]{\boldmath$\phi^2 \psi^2 H D^2 + \text{\bf h.c.}$}} 
\\ \hhline{~|~||--}
$\op{8}{\phi^2 G^2}{(1)}$ & $(\partial^\mu \phi\, \partial_\nu \phi)G_{\mu\rho}^A G^{A\nu \rho}$ & $\op{8}{\phi^2 \ell e H}{pr}$ & $(\partial_\mu \phi \, \partial^\mu \phi)(\overline \ell_p e_r H) $ 
\\ 
$\op{8}{\phi^2 G^2}{(2)}$ & $(\partial_\mu \phi \, \partial^\mu \phi)G_{\nu\rho}^A G^{A\nu\rho}$ & $\op{8}{\phi^2 qu H}{pr}$ & $(\partial_\mu \phi \, \partial^\mu \phi)(\overline q_p u_r \widetilde H) $ 
\\
$\op{8}{\phi^2 G^2}{(3)}$ & $(\partial_\mu \phi \, \partial^\mu \phi)G_{\nu\rho}^A \widetilde G^{A\nu\rho}$ & $\op{8}{\phi^2 qd H}{pr}$ & $(\partial_\mu \phi \, \partial^\mu \phi)(\overline q_p d_r H) $
\\ \hline
 \multicolumn{2}{c||}{\cellcolor{gray!20}\hyperref[par:phi4D4]{\boldmath$\phi^4 D^4$}} & \multicolumn{2}{c}{\cellcolor{gray!20}\hyperref[par:phi2H2D4]{\boldmath$\phi^2 H^2 D^4$}} 
 \\ \hline
$\op{8}{\phi^4}{}$ & $(\partial_\mu \phi \, \partial^\mu \phi)^2$ & $\op{8}{\phi^2 H^2}{(1)}$ & $(\partial_\mu \partial_\nu \phi\, \partial^\mu \partial^\nu \phi)(H^\dagger H)$ 
\\ \hhline{--||~|~}
\multicolumn{2}{c||}{\cellcolor{gray!20}\hyperref[par:phi2H4D2]{\boldmath$\phi^2 H^4 D^2$}} & $\op{8}{\phi^2 H^2}{(2)}$ & $(\partial^\mu \phi \, \partial^\nu \phi)(D_\mu H^\dagger D_\nu H)$ 
\\ \hhline{--||~|~}
$\op{8}{\phi^2 H^4}{}$ & $(\partial_\mu \phi \, \partial^\mu \phi)(H^\dagger H)^2$ & \multicolumn{2}{c}{} 
\\ \bottomrule
\end{tabular}
\caption{Dimension-8 ALP operators that conserve lepton and baryon numbers.
Operators with $+ \text{h.c.}$ have Hermitian conjugates.
The indices $p,r$ denote weak eigenstates. The labels of the operator classes contain a cross-reference to the corresponding unitarity bounds.\label{tab:dim8}}
\end{table}

\section{Unitarity bounds}
\label{unitarity_bounds}

An interesting framework for performing a detailed analysis of unitarity constraints is that of EFTs for ALPs. 
Indeed, the pNGB nature of the ALP forces its interactions with SM fields to be of derivative type and, therefore, to grow with the energy.
As a result, strong unitarity bounds are typically expected~\cite{Brivio:2021fog}.
Clearly, as the dimensionality of an operator increases, the energy growth of the interaction also increases, leading to more stringent unitarity constraints.
However, the larger the dimensionality of an operator, the smaller its expected phenomenological impact on physical observables. 
One could then question the approach itself of putting unitarity bounds on higher-dimensional operators if their effects are somehow expected to be subdominant in any case.
However, this is not necessarily the case, as was shown for instance in \cite{Bertuzzo:2023slg} for the production of a Higgs-ALP pair at a lepton collider: the impact of the dimension-7 operators $\partial_\mu \phi \,D^\mu \overline \ell H e$ and $\partial_\mu \phi \,\overline \ell H D_\mu e$ on the total cross section was there shown to be significantly larger than the one due to the dimension-5 operator $\phi \,\overline{\ell} H e$ in a wide region of the parameter space.
Motivated by these considerations, we stress the possible relevance of placing unitarity constraints on higher-dimensional ALP interactions, which constitutes the subject of this section.

\subsection{Dimension-5 and -6 operators}

The dimension-5 and -6 ALP operators are reported in Table~\ref{tab:dim5-6}.

\paragraph{$\phi X^2$ class\label{par:phiX2}.}

The strongest partial wave unitarity bound associated with the interaction $\coef{5}{\phi X^2}{}\,\phi\, X^{\mathscr A\mu\nu}\widetilde X_{\mu\nu}^{\mathscr A}$, where $X_{\mu\nu}^{\mathscr A}$ is the field strength of a generic gauge group $G$, $\widetilde X_{\mu\nu}^{\mathscr A} = \frac{1}{2}\epsilon_{\mu\nu\rho\sigma}X^{\mathscr A \rho\sigma}$ (with $\epsilon^{0123} = 1$), and $\mathscr A\in\{1,\dotsc,d(G)\}$ is an adjoint index, is
\begin{equation}\label{eq:phiX2_unibounds}
    \sqrt{s}\abs{\coef{5}{\phi X^2}{}} \le \min\left\{ \sqrt{\frac{4\pi}{2\,d(G)-1}},\sqrt{\frac{8\pi}{1+\sqrt{32\,d(G)+1}}}\right\}
\end{equation}
where $d(G)$ is the dimension of $G$.
This bound is derived from the $J=0$ partial wave coefficients of the amplitudes for $XX \to XX$ and $XX \to \phi \phi$, constructed with a double insertion of the operator $\op{5}{\phi X^2}{} = \phi\, X^{\mathscr A\mu\nu}\widetilde X_{\mu\nu}^{\mathscr A}$; see Appendix~\ref{app:amplitudes5e6} for their expressions.

This implies that the Wilson coefficients associated with the dimension-5 bosonic operators in Table~\ref{tab:dim5-6} are subject to the following individual constraints:\footnote{In Ref.~\cite{Brivio:2021fog} it is claimed that, for $\sqrt{s} \geq 260$~GeV, the strongest bound on $\coef{5}{\phi W^2}{}$ arises from the $W^+ W^- \to Z (\gamma) \phi$ processes as they are linearly dependent on $\coef{5}{\phi W^2}{}$ and scale with energy as $s^{3/2}$. By an explicit calculation of the associated scattering amplitudes, both in the broken and unbroken phases of the SM, we find instead a $s^{1/2}$ energy dependence, as it is also manifest by adopting the {\it{equivalent gauge}} of~\cite{Wulzer:2013mza}. As a result, the processes $XX \to XX$ and $XX \to \phi \phi$ set the most stringent bound on $\coef{5}{\phi W^2}{}$ at any energy scale.}
\begin{align}
    \sqrt{s}\abs{\coef{5}{\phi B^2}{}} &\le \sqrt{\frac{8\pi}{1+\sqrt{33}}} \approx 1.9\,,
    \\
    \sqrt{s}\abs{\coef{5}{\phi W^2}{}} &\le \sqrt{\frac{8\pi}{1+\sqrt{97}}}\approx 1.5\,,
    \\
    \sqrt{s}\abs{\coef{5}{\phi G^2}{}} &\le \sqrt{\frac{4\pi}{15}} \approx 0.92\,.
\end{align}

For example, the bound for $\coef{5}{\phi B^2}{}$ is obtained considering the scattering matrix
\begin{equation}
    \mathcal A_{i \to f} = 
    \begin{pmatrix}
        -4\,\coef{5}{\phi B^2}{2}\,\frac{\agl{3}{4}^2\sqr{2}{1}}{\agl{1}{2}} & 0 & 8\,\coef{5}{\phi B^2}{2}\,\sqr{2}{1}^2
        \\
        0 & -4\,\coef{5}{\phi B^2}{2}\,\frac{\agl{1}{2}^2\sqr{4}{3}}{\agl{3}{4}} & 8\,\coef{5}{\phi B^2}{2}\,\agl{1}{2}^2
        \\
        8\,\coef{5}{\phi B^2}{2}\,\agl{3}{4}^2 & 8\,\coef{5}{\phi B^2}{2}\,\sqr{4}{3}^2 & 0
    \end{pmatrix}
\end{equation}
where $i,f\in\{B^-B^-,B^+B^+,\phi \phi\}$.
The associated $J=0$ partial wave scattering matrix, after including the $1/\sqrt{2}$ factor for identical particle pairs in the initial and final states, is
\begin{equation}
    a^0_{i \to f} = \frac{s}{4\pi}\,\coef{5}{\phi B^2}{2}
    \begin{pmatrix}
        -1 & 0 & 2 \\
        0 & -1 & 2 \\
        2 & 2 & 0
    \end{pmatrix}\,,
\end{equation}
whose eigenvalues are $\frac{-s}{8\pi}(1\pm \sqrt{33})\,\coef{5}{\phi B^2}{2}$ and $\frac{-s}{4\pi}\,\coef{5}{\phi B^2}{2}$.
In this case, since the gauge bosons are Abelian, we have $\mathcal A (B^{\pm}B^{\pm} \to B^{\mp}B^{\mp}) = 0$.
On the other hand, if the gauge bosons are non-Abelian, the associated amplitude for the same helicity configuration no longer vanishes and contributes to having $\frac{-s}{4\pi}(2\,d(G)-1)\,\coef{5}{\phi X^2}{2}$ as an eigenvalue, which corresponds to the first term on the right-hand side of Eq.~\eqref{eq:phiX2_unibounds}.

\begin{figure}[h!tb]
    \centering
    \includegraphics[scale = 0.92]{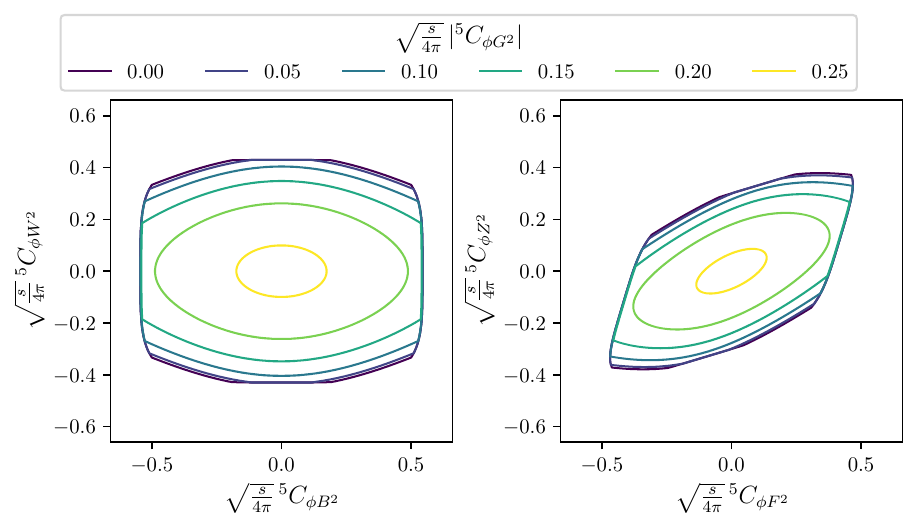}
    \caption{Plots of the boundaries of the parameter space allowed by the partial wave unitarity bounds in Eq.~\eqref{eq:PWUB_d5_phiX2} in the $\coef{5}{\phi B^2}{}$--$\coef{5}{\phi W^2}{}$ plane (\textit{left panel}) and in the $\coef{5}{\phi F^2}{}$--$\coef{5}{\phi Z^2}{}$ plane (\textit{right panel}) for different values of $|\coef{5}{\phi G^2}{}|$. The Wilson coefficients $\coef{5}{\phi F^2}{}$ and $\coef{5}{\phi Z^2}{}$ are defined as $\coef{5}{\phi F^2}{} = c^2_W \, \coef{5}{\phi B^2}{} + s^2_W\,\coef{5}{\phi W^2}{}$ and $\coef{5}{\phi Z^2}{} = s^2_W \, \coef{5}{\phi B^2}{} + c^2_W\,\coef{5}{\phi W^2}{}$ (where $s_W$ and $c_W$ are the sine and cosine of the weak mixing angle) and mediate the interactions of the ALP with photons and $Z$ bosons, respectively: $\mathcal L \supset \coef{5}{\phi F^2}{} \, \phi \, F_{\mu\nu}\widetilde F^{\mu\nu} + \coef{5}{\phi Z^2}{} \, \phi \, Z_{\mu\nu}\widetilde Z^{\mu\nu}$.}
    \label{fig:d5phiX2}
\end{figure}

Stronger bounds can be obtained by considering also scattering processes where these three Wilson coefficients are simultaneously non-vanishing.
In fact, in that case, the amplitudes for $BB \to WW$, $WW \to GG$, and $BB \to GG$ are each proportional to the product of the two different Wilson coefficients.
As a result, if we include them, the bounds on $\coef{5}{\phi B^2}{}$, $\coef{5}{\phi W^2}{}$, and $\coef{5}{\phi G^2}{}$ become correlated and more stringent than when considered independently.
The largest eigenvalues of the entire partial wave scattering matrix with $J=0$ are given by the roots $x_i$ of the cubic and quartic polynomials
\begin{equation}
    p(x) = a_0 + a_1 x+a_2x^2+x^3\,, \qquad
    q(x) = b_0 + b_1 x + b_2 x^2 + b_3 x^3 + x^4\,,
\end{equation}
with
\begin{align}
    a_0 &= 23\, \coef{5}{\phi B^2}{2}\,\coef{5}{\phi W^2}{2}\,\coef{5}{\phi G^2}{2}\,R^6 \,,\\
    a_1&= - \left(17\, \coef{5}{\phi B^2}{2}\,\coef{5}{\phi G^2}{2}+7\,\coef{5}{\phi B^2}{2}\,\coef{5}{\phi W^2}{2}+21\,\coef{5}{\phi W^2}{2}\,\coef{5}{\phi G^2}{2}\right)R^4\,,\\
    a_2&= \left(\coef{5}{\phi B^2}{2} + 5\,\coef{5}{\phi W^2}{2} + 15\,\coef{5}{\phi G^2}{2}\right) R^2\,,
\end{align}
and
\begin{align}
    b_0 &= -8\, \coef{5}{\phi B^2}{2}\,\coef{5}{\phi W^2}{2}\,\coef{5}{\phi G^2}{2} \left(\coef{5}{\phi B^2}{2}+3\,\coef{5}{\phi W^2}{2}+8\,\coef{5}{\phi G^2}{2}\right) R^8\,,
    \\
    b_1 &= 
    \Big(\coef{5}{\phi B^2}{2}\,\coef{5}{\phi W^2}{2}\,\coef{5}{\phi G^2}{2} -8\,\coef{5}{\phi B^2}{4}\,\coef{5}{\phi W^2}{2}
    -24\,\coef{5}{\phi B^2}{2}\,\coef{5}{\phi W^2}{4}
    -8\,\coef{5}{\phi B^2}{4}\,\coef{5}{\phi G^2}{2}
    \nonumber \\&\quad -24\,\coef{5}{\phi W^2}{4}\,\coef{5}{\phi G^2}{2}
    -64\, \coef{5}{\phi B^2}{2}\,\coef{5}{\phi G^2}{4}
    -64\,\coef{5}{\phi W^2}{2}\,\coef{5}{\phi G^2}{4}\Big) \, R^6\,,
    \\
    b_2 &= 
    \left[
    \coef{5}{\phi B^2}{2}\,\coef{5}{\phi W^2}{2}
    +\coef{5}{\phi B^2}{2}\,\coef{5}{\phi G^2}{2}
    +\coef{5}{\phi W^2}{2}\,\coef{5}{\phi G^2}{2} -8\left( \coef{5}{\phi B^2}{4}+3\,\coef{5}{\phi W^2}{4}+8\,\coef{5}{\phi G^2}{4} \right)\right]R^4\,,
    \\
    b_3 &= \left(\coef{5}{\phi B^2}{2} + \coef{5}{\phi W^2}{2} + \coef{5}{\phi G^2}{2}\right)R^2\,,
\end{align}
where $R=\sqrt{s/(4\pi)}$.
The corresponding unitarity bounds are then found by imposing $\abs{x_i} \le 1$ on each root.
Using the Jury stability criterion~\cite{Schur1918} (see also \cite{grove2004periodicities}, Theorems 1.4 and 1.5), we can avoid the explicit calculation of the roots of $p(x)$ and $q(x)$. 
In fact, $\abs{x_i} \le 1$ holds for each root of $p(x)$ and $q(x)$ if and only if all the following inequalities are satisfied, respectively:
\begin{align}
\label{eq:PWUB_d5_phiX2}
    \begin{dcases}
        \abs{a_2+a_0} \le 1 + a_1\,,\\
        \abs{a_2 - 3 a_0 } \le 3- a_1\,,\\
        a_0^2 + a_1 - a_0 a_2 \le 1\,,
    \end{dcases}
    &&
    \begin{dcases}
        \abs{b_1 + b_3} \le 1+b_0+b_2\,,\\
        \abs{b_1 - b_3} \le 2(1-b_0)\,,\\
        b_2 - 3 b_0 \le 3\,,\\
        b_1^2 + b_2 + b_0(1+b_0+b_0 b_2+b_3^2) \le 1 + b_0^3 + 2 b_0 b_2 + (1+b_0) b_1 b_3\,.
    \end{dcases}
\end{align}
The corresponding plots, where we take the intersection of the two regions (which is equivalent to taking the most stringent bound), are shown in Fig.~\ref{fig:d5phiX2}.

\paragraph{$\phi \psi^2 H$ class\label{par:phipsi2H}.}

The individual bounds for this class of Wilson coefficients are
\begin{gather}
    s \sum_{p,r}\abs{\coef{5}{\phi \ell e H}{pr}}^2 \le 64\pi^2\,,
    \qquad
    s \sum_{p,r}\abs{\coef{5}{\phi qu H}{pr}}^2 \le \frac{64}{3}\pi^2\,,
    \qquad
    s \sum_{p,r}\abs{\coef{5}{\phi qd H}{pr}}^2 \le \frac{64}{3}\pi^2\,,
\end{gather}
while the combined bound is
\begin{equation}
    s \sum_{p,r}\left(\abs{\coef{5}{\phi \ell e H}{pr}}^2 + 3\abs{\coef{5}{\phi qu H}{pr}}^2+3\abs{\coef{5}{\phi qd H}{pr}}^2 \right) \le 64\pi^2\,.
\end{equation}
These are derived from the $J=0$ partial wave coefficients of the amplitudes for $\phi H \to \psi_p \overline \psi_r$.

\paragraph{$\phi^2 H^2 D^2$ class\label{par:phi2H2D2}.}

The bound for this Wilson coefficient is
\begin{equation}
    s \abs{\coef{6}{\phi ^2 H^2}{}} \le 8\pi\,,
\end{equation}
which is derived from the $J=0$ partial wave coefficients of the amplitudes for $\phi \phi \to H\overline H$.

\begin{figure}[h!tb]
    \centering
    \includegraphics[scale = 0.92]{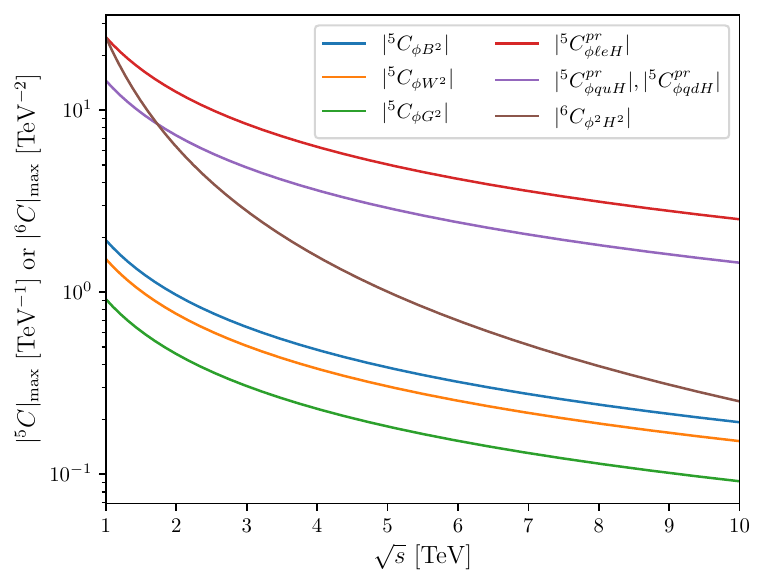}
    \caption{Summary of the marginalized partial wave unitarity bounds on the Wilson coefficients associated with the dimension-5 and -6 ALP operators.
    }
    \label{fig:d56summary}
\end{figure}

In Fig.~\ref{fig:d56summary}, we summarize the marginalized partial wave unitarity bounds on the Wilson coefficients associated with the dimension-5 and -6 ALP operators.

\subsection{Dimension-7 operators}

The dimension-7 ALP operators are reported in Table~\ref{tab:dim7}.

\paragraph{$\phi \psi^2 H D^2$ class\label{par:phipsi2HD2}.}

\begin{figure}[h!tb]
    \centering
    \includegraphics[scale = 0.92]{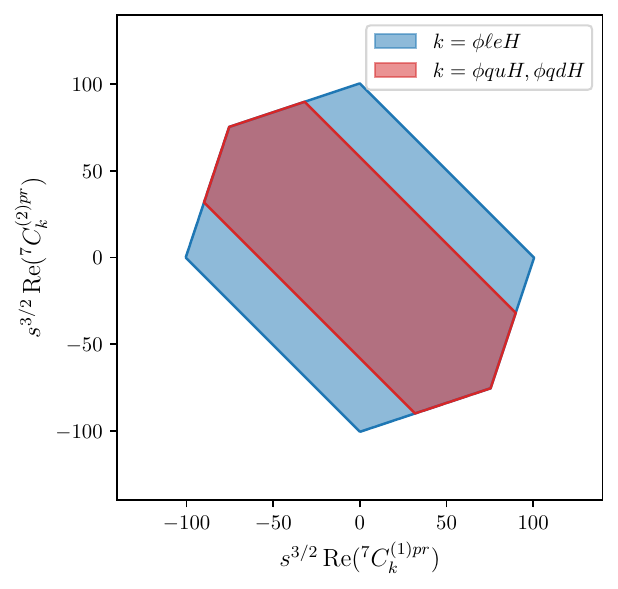}
    \caption{Parameter space allowed by partial wave unitarity bounds in Eqs.~\eqref{eq:PWUB_d7_phipsi2HD2_1}--\eqref{eq:PWUB_d7_phipsi2HD2_2} associated with the dimension-7 operators $\op{7}{k}{(1)pr}$ and $\op{7}{k}{(2)pr}$, with $k=\phi \ell e H,\phi qu H,\phi qd H$ and $p,r=1,2,3$. The imaginary parts of the Wilson coefficients are taken to be zero.}
    \label{fig:d7phipsi2H}
\end{figure}

The individual bounds for this class of Wilson coefficients are
\begin{gather}
    s^3 \sum_{p,r}\abs{\coef{7}{\phi \ell e H}{(1)pr} + \coef{7}{\phi \ell e H}{(2)pr}}^2 \le 1024\pi^2\,, \label{eq:PWUB_d7_phipsi2HD2_1}
    \\
    s^{3/2} \abs{3\,\coef{7}{\phi \ell e H}{(1)pr} - \coef{7}{\phi \ell e H}{(2)pr}} \le 96\pi\,,
    \qquad
    s^{3/2} \abs{\coef{7}{\phi \ell e H}{(1)pr} - 3\, \coef{7}{\phi \ell e H}{(2)pr}} \le 96\pi\,,
    \\
    s^3 \sum_{p,r}\abs{\coef{7}{\phi qu H}{(1)pr} + \coef{7}{\phi qu H}{(2)pr}}^2 \le \frac{1024}{3}\pi^2\,,
    \\
    s^{3/2} \abs{3\,\coef{7}{\phi qu H}{(1)pr} - \coef{7}{\phi qu H}{(2)pr}} \le 96\pi\,,
    \qquad
    s^{3/2} \abs{\coef{7}{\phi qu H}{(1)pr} - 3\, \coef{7}{\phi qu H}{(2)pr}} \le 96\pi\,,
    \\
    s^3 \sum_{p,r}\abs{\coef{7}{\phi qd H}{(1)pr} + \coef{7}{\phi qd H}{(2)pr}}^2 \le \frac{1024}{3}\pi^2\,,
    \\
    s^{3/2} \abs{3\,\coef{7}{\phi qd H}{(1)pr} - \coef{7}{\phi qd H}{(2)pr}} \le 96\pi\,,
    \qquad
    s^{3/2} \abs{\coef{7}{\phi qd H}{(1)pr} - 3\, \coef{7}{\phi qd H}{(2)pr}} \le 96\pi\,, \label{eq:PWUB_d7_phipsi2HD2_2}
\end{gather}
while the combined bound is
\begin{equation}
    s^3
    \sum_{p,r}\left(
    \abs{\coef{7}{\phi \ell e H}{(1)pr} + \coef{7}{\phi \ell e H}{(2)pr}}^2
    + 3
    \abs{\coef{7}{\phi qu H}{(1)pr} + \coef{7}{\phi qu H}{(2)pr}}^2
    + 3
    \abs{\coef{7}{\phi qd H}{(1)pr} + \coef{7}{\phi qd H}{(2)pr}}^2
    \right)
     \le 1024\pi^2\,.
\end{equation}
These are derived from the $J=0$ partial wave coefficients of the amplitudes for $\phi H \to \psi_p \overline \psi_r$ and the $J=1/2$ partial wave coefficients of the amplitudes for $\phi \psi_p \to H \psi_r$.
The bounds are reported in Fig.~\ref{fig:d7phipsi2H}.

\paragraph{$\phi \psi^2 X D$ class\label{par:phipsi2XD}.}

The bounds for the Wilson coefficients belonging to this class are
\begin{align}
    s^{3} \sum_{p,r} \abs{\coef{7}{\phi \ell^2 B}{(1)pr}+ i \,\coef{7}{\phi \ell^2 B}{(2)pr}}^2 &\le 288\pi^2\,,
     &
    s^{3} \sum_{p,r} \abs{\coef{7}{\phi e^2 B}{(1)pr}+ i \,\coef{7}{\phi e^2 B}{(2)pr}}^2 &\le 576\pi^2\,,    
    \\
    s^{3} \sum_{p,r} \abs{\coef{7}{\phi q^2 B}{(1)pr}+ i \,\coef{7}{\phi q^2 B}{(2)pr}}^2 &\le 96 \pi^2\,,
    &
    s^{3} \sum_{p,r} \abs{\coef{7}{\phi u^2 B}{(1)pr}+ i \,\coef{7}{\phi u^2 B}{(2)pr}}^2 &\le 192 \pi^2\,,
    \\
    s^{3} \sum_{p,r} \abs{\coef{7}{\phi d^2 B}{(1)pr}+ i \,\coef{7}{\phi d^2 B}{(2)pr}}^2 &\le 192 \pi^2\,,
    &
    s^{3} \sum_{p,r} \abs{\coef{7}{\phi \ell^2 W}{(1)pr}+ i \,\coef{7}{\phi \ell^2 W}{(2)pr}}^2 &\le 288\pi^2\,,
    \\
    s^{3} \sum_{p,r} \abs{\coef{7}{\phi q^2 W}{(1)pr}+ i \,\coef{7}{\phi q^2 W}{(2)pr}}^2 &\le 96 \pi^2\,,
    &
    s^{3} \sum_{p,r} \abs{\coef{7}{\phi q^2 G}{(1)pr}+ i \,\coef{7}{\phi q^2 G}{(2)pr}}^2 &\le 144 \pi^2\,, 
    \\
    s^{3} \sum_{p,r} \abs{\coef{7}{\phi u^2 G}{(1)pr}+ i \,\coef{7}{\phi u^2 G}{(2)pr}}^2 &\le 288 \pi^2\,,
    & 
    s^{3} \sum_{p,r} \abs{\coef{7}{\phi d^2 G}{(1)pr}+ i \,\coef{7}{\phi d^2 G}{(2)pr}}^2 &\le 288 \pi^2\,.
\end{align}
These are derived from the $J=1$ partial wave coefficients of the amplitudes for $\phi X \to \psi_p \overline \psi_r$.

\paragraph{$\phi X H^2 D^2$ class\label{par:phiXH2D2}.}

The bounds for the Wilson coefficients belonging to this class are
\begin{equation}
s^{3/2}\abs{\coef{7}{\phi B H^2}{(1)}+ i \,\coef{7}{\phi B H^2}{(2)}} \le 24\pi\,,
\qquad
s^{3/2}\abs{\coef{7}{\phi W H^2}{(1)}+ i \,\coef{7}{\phi W H^2}{(2)}} \le 24\pi\,.
\end{equation}
These are derived from the $J=1$ partial wave coefficients of the amplitudes for $\phi X \to H \overline H$.

\paragraph{$\phi H^4 D^2$ class\label{par:phiH4D2}.}
From the $J=0$ partial wave coefficients of the amplitudes for $\phi H \to HH\overline H$ we obtain
\begin{equation}
    s^{3/2} \abs{\coef{7}{\phi H^4}{}} \le 16\sqrt{6}\pi^2\,.
\end{equation}

\paragraph{$\phi \psi^2 H^2 D$ class\label{par:phipsi2H2D}.}
The individual bounds for the Wilson coefficients belonging to this class are
\begin{gather}
    s^3 \sum_{p,r}\abs{\coef{7}{\phi \ell^2 H^2}{(1)pr}}^2 \le (32\sqrt{6}\pi^2)^2 \,,
    \qquad 
    s^3 \sum_{p,r}\abs{\coef{7}{\phi \ell^2 H^2}{(2)pr}}^2 \le (32\sqrt{6}\pi^2)^2 \,,
    \\
    s^3 \sum_{p,r}\abs{\coef{7}{\phi q^2 H^2}{(1)pr}}^2 \le (32\sqrt{2}\pi^2)^2 \,,
    \qquad 
    s^3 \sum_{p,r}\abs{\coef{7}{\phi q^2 H^2}{(2)pr}}^2 \le (32\sqrt{2}\pi^2)^2  \,,
    \\
    s^3 \sum_{p,r}\abs{\coef{7}{\phi e^2 H^2}{pr}}^2 \le (64\sqrt{3}\pi^2)^2 \,,
    \;\,\, 
    s^3 \sum_{p,r}\abs{\coef{7}{\phi u^2 H^2}{pr}}^2 \le (64\pi^2)^2  \,,
    \;\,\,
    s^3 \sum_{p,r}\abs{\coef{7}{\phi d^2 H^2}{pr}}^2 \le (64\pi^2)^2  \,.
\end{gather}
They are derived from the $J=0$ partial wave coefficients of the amplitudes for $H \overline H \to \phi \psi_p \overline \psi_{r}$.
The combined bounds are
\begin{gather}
s^3 \sum_{p,r}\left(\abs{\coef{7}{\phi \ell^2 H^2}{(2)pr}}^2+3\abs{\coef{7}{\phi q^2 H^2}{(2)pr}}^2\right) \le (32\sqrt{6}\pi^2)^2\,,
\\
    s^3 \sum_{p,r}\left(\abs{\coef{7}{\phi e^2 H^2}{pr}}^2 + 3\abs{\coef{7}{\phi u^2 H^2}{pr}}^2+3\abs{\coef{7}{\phi d^2 H^2}{pr}}^2+2\abs{\coef{7}{\phi \ell^2 H^2}{(1)pr}}^2+6\abs{\coef{7}{\phi q^2 H^2}{(1)pr}}^2\right) \le (64\sqrt{3}\pi^2)^2\,.
\end{gather}

\begin{figure}[h!tb]
    \centering
    \includegraphics[scale = 0.92]{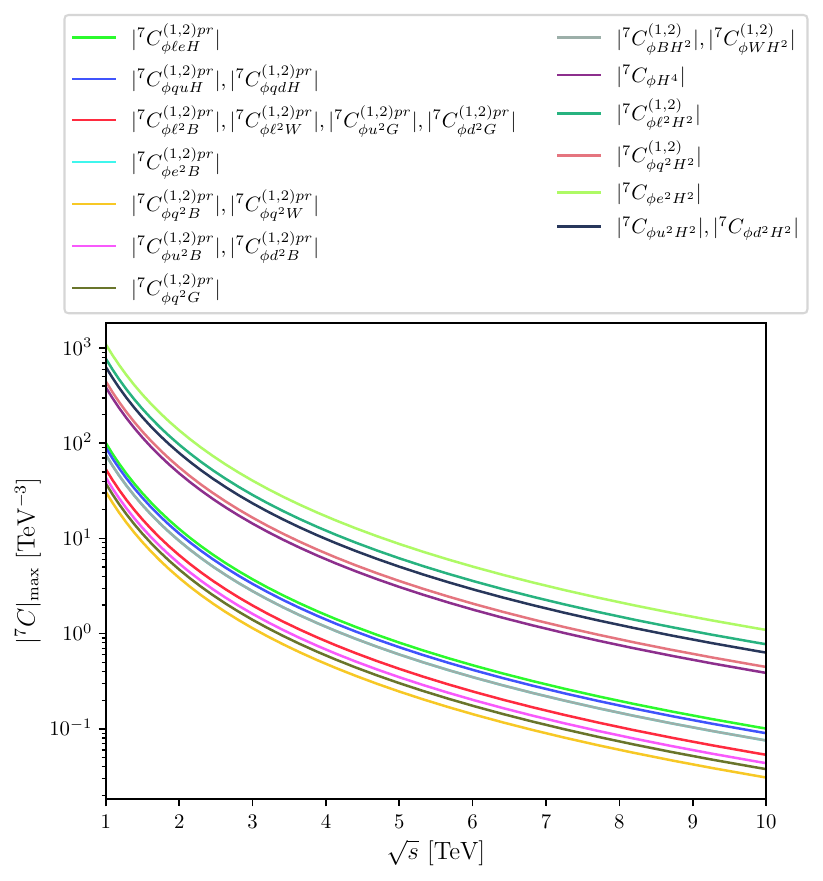}
    \caption{Summary of the marginalized partial wave unitarity bounds on the Wilson coefficients associated with the dimension-7 ALP operators.
    }
    \label{fig:d7summary}
\end{figure}

In Fig.~\ref{fig:d7summary}, we summarize the marginalized partial wave unitarity bounds on the Wilson coefficients associated with the dimension-7 ALP operators.

\subsection{Dimension-8 operators}

The dimension-8 ALP operators that conserve lepton and baryon numbers are reported in Table~\ref{tab:dim8}.

\paragraph{$\phi^4 D^4$ class\label{par:phi4D4}.}
The strongest partial wave unitarity bound for this Wilson coefficient is
\begin{equation}
    s^2 \abs{\coef{8}{\phi^4}{}} \le \frac{24}{5}\pi
\end{equation}
and is derived from the $J=0$ partial wave coefficient of the amplitude for $\phi \phi \to \phi \phi$.

\paragraph{$\phi^2 H^4 D^2$ class\label{par:phi2H4D2}.}
The strongest partial wave unitarity bound for this Wilson coefficient is
\begin{equation}
    s^2 \abs{\coef{8}{\phi^2 H^4}{}} \le 128\sqrt{2}\pi^3
\end{equation}
and is derived from the $J=0$ partial wave coefficients of the amplitudes for $\phi \phi \to H\overline HH\overline H$.

\paragraph{$\phi^2 H^2 D^4$ class\label{par:phi2H2D4}.}
The partial wave unitarity bounds for this class of Wilson coefficients are
\begin{equation}
    \label{eq:d8-uni_phi2H2}
    s^2 \abs{3\, \coef{8}{\phi^2 H^2}{(1)}+\coef{8}{\phi^2 H^2}{(2)}} \le 48 \pi\,, 
    \qquad 
    s^2  \abs{\coef{8}{\phi^2 H^2}{(1)}+2\, \coef{8}{\phi^2 H^2}{(2)}} \le 48 \pi\,,
\end{equation}
which are derived from the $J=0$ partial wave coefficients of the amplitudes for $\phi \phi \to H\overline H$ and $\phi H \to \phi H$.
They are shown in Fig.~\ref{fig:d8phi2H2}.

\paragraph{$\phi^2 X^2 D^2$ class\label{par:phi2X2D2}.}
The individual bounds for this class of Wilson coefficients are
\begin{gather}\label{eq:PWUB_d8_phi2X2_1}
s^2 \sqrt{\left(\coef{8}{\phi^2 B^2}{(1)}+4\,\coef{8}{\phi^2 B^2}{(2)}\right)^2+\left(4\,\coef{8}{\phi^2 B^2}{(3)}\right)^2} \le 16 \sqrt{2}\pi\,, \\
s^2 \left[8\abs{\coef{8}{\phi^2 B^2}{(1)}}+3\sqrt{\left(\coef{8}{\phi^2 B^2}{(1)}+4\,\coef{8}{\phi^2 B^2}{(2)}\right)^2+\left(4\,\coef{8}{\phi^2 B^2}{(3)}\right)^2}\right]\le 192 \pi\,,
\\
s^2 \sqrt{\left(\coef{8}{\phi^2 W^2}{(1)}+4\,\coef{8}{\phi^2 W^2}{(2)}\right)^2+\left(4\,\coef{8}{\phi^2 W^2}{(3)}\right)^2} \le 16 \sqrt{\frac{2}{3}}\pi\,, 
\\
s^2 \left[8\abs{\coef{8}{\phi^2 W^2}{(1)}}+3\sqrt{\left(\coef{8}{\phi^2 W^2}{(1)}+4\,\coef{8}{\phi^2 W^2}{(2)}\right)^2+\left(4\,\coef{8}{\phi^2 W^2}{(3)}\right)^2}\right]\le 192 \pi\,,
\\
s^2 \sqrt{\left(\coef{8}{\phi^2 G^2}{(1)}+4\,\coef{8}{\phi^2 G^2}{(2)}\right)^2+\left(4\,\coef{8}{\phi^2 G^2}{(3)}\right)^2} \le 8\pi\,,
\\
s^2 \left[8\abs{\coef{8}{\phi^2 G^2}{(1)}}+3\sqrt{\left(\coef{8}{\phi^2 G^2}{(1)}+4\,\coef{8}{\phi^2 G^2}{(2)}\right)^2+\left(4\,\coef{8}{\phi^2 G^2}{(3)}\right)^2}\right]\le 192 \pi\,,
\label{eq:PWUB_d8_phi2X2_2}
\end{gather}
while the combined one is
\begin{align}
    s^4\bigg[&\left(\coef{8}{\phi^2 B^2}{(1)}+4\,\coef{8}{\phi^2 B^2}{(2)}\right)^2+\left(4\,\coef{8}{\phi^2 B^2}{(3)}\right)^2 + 3 \left(\coef{8}{\phi^2 W^2}{(1)}+4\,\coef{8}{\phi^2 W^2}{(2)}\right)^2+3\left(4\,\coef{8}{\phi^2 W^2}{(3)}\right)^2\nonumber \\
    &+ 8\left(\coef{8}{\phi^2 G^2}{(1)}+4\,\coef{8}{\phi^2 G^2}{(2)}\right)^2+8\left(4\,\coef{8}{\phi^2 G^2}{(3)}\right)^2\bigg]\le (16\sqrt{2}\pi)^2\,.
\end{align}
These bounds are derived from the $J=0$ partial wave coefficients of the amplitudes for $\phi \phi \to XX$ and the $J=1$ partial wave coefficients of the amplitudes for $\phi X \to \phi X$.
They are shown in Fig.~\ref{fig:d8phi2B2}.

\paragraph{$\phi^2 \psi^2 D^3$ class\label{par:phi2psi2D3}.}

The strongest individual bounds are
\begin{equation}
    s^2 \abs{\coef{8}{\phi^2 \psi^2}{pr}} \le \frac{96}{5}\pi 
    \qquad
    \text{with~}
    \psi = \ell,e,q,u,d
\end{equation}
and are derived from the $J=1/2$ partial wave coefficients of the amplitudes for $\phi \psi_p \to \phi \psi_r$.
The collective bound is
\begin{equation}
    s^4 \sum_{p,r}\bigg(
    \abs{\coef{8}{\phi^2 e^2}{pr}}^2 
    +2\abs{\coef{8}{\phi^2 \ell^2}{pr}}^2
    + 3 \abs{\coef{8}{\phi^2 u^2}{pr}}^2 
    + 3 \abs{\coef{8}{\phi^2 d^2}{pr}}^2
    + 6 \abs{\coef{8}{\phi^2 q^2}{pr}}^2 
    \bigg) \le (160\sqrt{3}\pi)^2
\end{equation}
and is derived from the $J=2$ partial wave coefficients of the amplitudes for $\phi \phi \to \psi_p \overline \psi_r$.

\paragraph{$\phi^2 \psi^2 H D^2$ class\label{par:phi2psi2HD2}.}
The individual bounds for this class of Wilson coefficients are
\begin{gather}
    s^4 \sum_{p,r}\abs{\coef{8}{\phi^2 \ell e H}{pr}}^2 \le (32\sqrt{3}\pi^2)^2\,,\\
    s^4 \sum_{p,r}\abs{\coef{8}{\phi^2 qu H}{pr}}^2 \le (32\pi^2)^2\,,
    \qquad 
    s^4 \sum_{p,r}\abs{\coef{8}{\phi^2 qd H}{pr}}^2 \le (32\pi^2)^2\,.
\end{gather}
The combined one is
\begin{equation}
    s^4 \sum_{p,r} \left(\abs{\coef{8}{\phi^2 \ell e H}{pr}}^2 + 3 \abs{\coef{8}{\phi^2 qu H}{pr}}^2 + 3 \abs{\coef{8}{\phi^2 qd H}{pr}}^2\right) \le (32\sqrt{3}\pi^2)^2\,.
\end{equation}
These are derived from the $J=0$ partial wave coefficients of the amplitudes for $\phi \phi \to \psi_p \overline \psi_r H$.

\begin{figure}[h!tb]
    \centering
    \includegraphics[scale = 0.92]{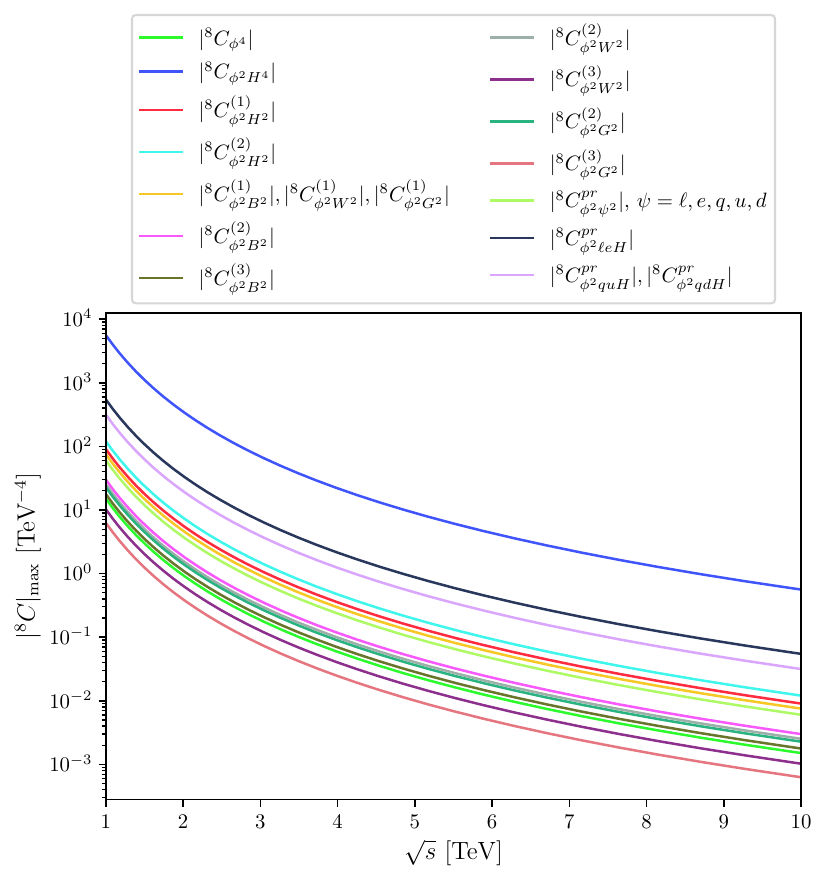}
    \caption{Summary of the marginalized partial wave unitarity bounds on the Wilson coefficients associated with the dimension-8 ALP operators.
    }
    \label{fig:d8summary}
\end{figure}

In Fig.~\ref{fig:d8summary}, we summarize the marginalized partial wave unitarity bounds on the Wilson coefficients associated with the dimension-8 ALP operators.

\subsection{Weak-violating ALP-lepton interactions}
The EFT assumed so far was invariant under the electroweak gauge group. If we give up this assumption, ALP-lepton interactions are described by the dimension-5 
Lagrangian~\cite{Altmannshofer:2022ckw}:
\begin{equation}
\mathcal{L} = \partial_\mu\phi \frac{\overline g_{ll }}{2m_l } \overline l  \gamma^\mu l  + \partial_\mu\phi \frac{g_{ll}}{2m_l} \overline l  \gamma^\mu \gamma_5 l  + \partial_\mu\phi \frac{g_{\nu _l }}{2m_l } \overline \nu_l \gamma^\mu P_L \nu_l \,,\qquad\qquad l = e,\, \mu,\, \tau,
\end{equation} 
where the $SU(2)_L$ invariance is recovered for $g_{\nu_l} = \overline g_{ll} - g_{ll}$. The implications of the above interactions are better understood by integrating the Lagrangian by parts~\cite{Altmannshofer:2022ckw}:
\begin{align}
-\mathcal{L} &= g_{ll} (\overline l i\gamma_5 l)\, \phi
+ \frac{ e ^2 }{ 16 \pi ^2 m _l } \bigg[ \frac{ \overline{g} _{ll} - 
g _{ll} + g _{\nu_l} }{ 4 s _W ^2} W ^+ _{\mu\nu} \widetilde W ^{ - , \mu \nu } \notag  + \frac{\overline{g} _{ll} - g _{ ll} ( 1 - 4  s _W ^2 ) }{ 2 c _W s _W } F _{\mu\nu} \widetilde{Z} ^{\mu\nu}
\\
&- g _{ll}  F _{\mu\nu} \widetilde{F} ^{\mu\nu} + \notag \frac{ \overline{g} _{ ll} ( 1 - 4 s _W ^2 ) - g _{ll} ( 1 - 4 s _W ^2 + 8 s _W ^4 )  + g _{\nu _l } }{ 8 s _W ^2 c _W ^2 } Z _{\mu\nu} \widetilde{Z} ^{\mu\nu} \bigg] \,\phi \notag
\\
& + \frac{ig}{2\sqrt{2} m _l }(g_{ll} - \overline g_{ll} + g_{\nu _l }) (\overline l \gamma^\mu P _L \nu_l) W_\mu^- \,\phi ~+~\text{h.c.}\,, 
\label{eq:weak_violating}
\end{align}
where $s_W $ is the sine of the weak mixing angle.
All terms in Eq.~\eqref{eq:weak_violating} are generated 
in the $SU(2)_L$ symmetric limit, except for the last term, which is a genuine electroweak-violating interaction.
As shown in Ref.~\cite{Altmannshofer:2022ckw}, this term generates
an $ ({\rm energy} / m_l )$ enhancement in various processes such as charged meson decays and $W$ boson decays.

\begin{figure}[h!tb]
    \centering
    \includegraphics[scale = 0.92]{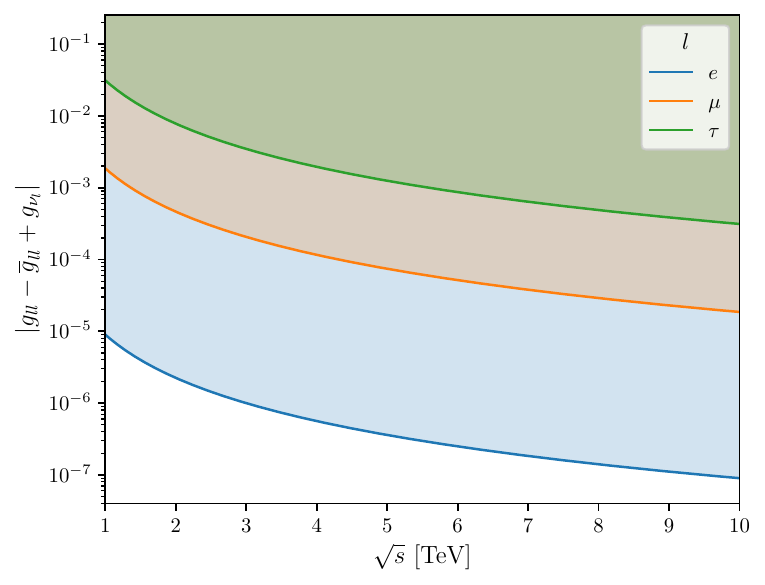}
    \caption{Excluded values of the weak-violating combinations $\abs{g_{ll}-\overline{g}_{ll}+g_{\nu_l}}$ as function of $\sqrt{s}$ by imposing the partial wave unitarity bounds of Eq.~\eqref{eq:WV_bounds}.}
    \label{fig:WV}
\end{figure}

The most stringent unitarity bound for the weak-violating combination of coefficients $g_{ll}-\overline g_{ll} + g_{\nu_l}$ is
\begin{equation}\label{eq:WV_bounds}
    \frac{s}{m_W m_l} g \abs{g_{ll}-\overline g_{ll} + g_{\nu_l}} \le 32\sqrt{2}\pi\,,
\end{equation}
which is derived from the $J=1/2$ partial wave coefficient of the amplitude for $\phi l \to W \nu_l$.
In particular, we have the following bounds
\begin{equation}
    \abs{g_{ll}-\overline g_{ll} + g_{\nu_l}} \lesssim 10^{-5} \left(\frac{m_l}{m_e}\right)
    \left(\frac{\text{TeV}}{\sqrt{s}}\right)^2\,,
\end{equation}
which are reported in Fig.~\ref{fig:WV}.

\section{Positivity bounds}
\label{sec:positivity}

Positivity bounds, stemming from the requirement that an EFT is the low-energy limit of a unitary, local, and causal quantum field theory~\cite{Adams:2006sv},
provide another set of conditions to infer the viable parameter space of consistent EFTs. 
Hereafter, we evaluate the complete set 
of positivity bounds for ALP EFTs, 
associated with dimension-8 operators,
discussing their complementarity and interplay with partial wave unitarity constraints~\cite{Bresciani:2025toe}.

Positivity bounds are derived from elastic $2 \to 2$ scattering processes whose amplitudes grow with even powers $n$ of $s$, with $n \ge 2$. 
By considering the scattering of a superposition of states $\ket{u} = \sum_i u_i \ket{i}$ and $\ket{v} = \sum_j v_j \ket{j}$
\begin{equation}
    \mathcal A (s,t) = \mathcal A(u,v \to u,v) = \sum_{i,j,k,l} u_i^* v_j^* u_k v_l \mathcal A(i,j\to k,l) \,,
\end{equation}
we can set in the forward limit \cite{Adams:2006sv,Bellazzini:2016xrt}
\begin{equation}
    \left.\dv[2]{}{s} \mathcal A(s,0)\right|_{s=0} \ge 0\,.
\end{equation}

For example, with regard to the operators $\op{8}{\phi^2 H^2}{(1)}$ and $\op{8}{\phi^2 H^2}{(2)}$, if we take 
\begin{equation}
    \ket{u} = \ket{\phi} + x \ket{\pi_a} \,, \qquad \ket{v} = \ket{\phi} + \ket{\pi_a}\,,
\end{equation}
where $\pi_a$ is any of the real components of the Higgs doublet
\begin{equation}
    H =\frac{1}{\sqrt{2}} \begin{pmatrix}
        \pi_1 + i\, \pi_3 \\ \pi_2 + i\, \pi_4 
    \end{pmatrix}\,,
\end{equation}
the amplitude is
\begin{align}
    \mathcal A (s,\theta) &= \mathcal A (u,v \to u,v) \nonumber\\&
    = \frac{s^2}{32}\bigg[
    \coef{8}{\phi^2 H^2}{(1)}\,(6 + 44x + 6x^{2})
  + \coef{8}{\phi^2 H^2}{(2)}\,(11+34x+11x^2)
  \nonumber  \\&\quad - 4 \left(2\,\coef{8}{\phi^2 H^2}{(1)} - \coef{8}{\phi^2 H^2}{(2)}\right)(1 - x)^{2}\cos\theta \nonumber  \\&\quad
  + \left(2\,\coef{8}{\phi^2 H^2}{(1)}\,(1 + x)^{2} + \coef{8}{\phi^2 H^2}{(2)}\,(1+6x+x^2)\right)\cos(2\theta)
\bigg]
\end{align}
and in the forward limit reduces to
\begin{equation}
    \mathcal A (s,0) = \frac{s^2}{2}\left[4\, \coef{8}{\phi^2 H^2}{(1)}\,x +  \coef{8}{\phi^2 H^2}{(2)}\, (1+x)^2\right]\,,
\end{equation}
leading to the bound in Eq.~\eqref{eq:d8-pos_phi2H2} by imposing $4\, \coef{8}{\phi^2 H^2}{(1)}\,x +  \coef{8}{\phi^2 H^2}{(2)} \,(1+x)^2 \ge 0$ for every $x \in \mathbb{R}$.

Another example is provided by the operators $\op{8}{\phi^2 B^2}{(1)}$, $\op{8}{\phi^2 B^2}{(2)}$, and $\op{8}{\phi^2 B^2}{(3)}$.
By parameterizing $\ket{u}$ and $\ket{v}$ as
\begin{equation}
    \ket{u} = \ket{\phi}+x_-\ket{B^-}+x_+\ket{B^+}\,,
    \qquad 
    \ket{v} = \ket{\phi}+y_-\ket{B^-}+y_+\ket{B^+}\,,
\end{equation}
and by using the amplitudes reported in Eqs.~\eqref{eq:ampl_d8_phi2B2(1)} and \eqref{eq:ampl_d8_phi2B2(2)},
the second derivative of $\mathcal A(s,\theta) = \mathcal A(u,v \to u,v)$ with respect to $s$ in the forward limit can be written as
\begin{equation}
    \left.\dv[2]{}{s}\mathcal A(s,0)\right|_{s=0} = \mathbf V\,\mathbf A\,\mathbf V^{T}\,,
\end{equation}
with
\begin{align}
    \mathbf V &=
    \begin{pmatrix}
        \Re x_- & \Im x_- & \Re x_+ & \Im x_+ & \Re y_- & \Im y_- & \Re y_+ & \Im y_+
    \end{pmatrix}\,,
    \\
    \mathbf A &=
    \begin{pmatrix}
-2\,\coef{8}{\phi^2 B^2}{(1)}\,\mathbb{I}_{4} & \mathbf B \\
\mathbf B & -2\,\coef{8}{\phi^2 B^2}{(1)}\,\mathbb{I}_{4}
\end{pmatrix}\,,
\end{align}
and
\begin{equation}
    \mathbf B=
    \begin{pmatrix}
\coef{8}{\phi^2 B^2}{(1)} + 4\,\coef{8}{\phi^2 B^2}{(2)} & -4\,\coef{8}{\phi^2 B^2}{(3)} & \coef{8}{\phi^2 B^2}{(1)} + 4\,\coef{8}{\phi^2 B^2}{(2)} & 4\,\coef{8}{\phi^2 B^2}{(3)} \\
-4\,\coef{8}{\phi^2 B^2}{(3)} & -\coef{8}{\phi^2 B^2}{(1)} - 4\,\coef{8}{\phi^2 B^2}{(2)} & -4\,\coef{8}{\phi^2 B^2}{(3)} & \coef{8}{\phi^2 B^2}{(1)} + 4\,\coef{8}{\phi^2 B^2}{(2)} \\
\coef{8}{\phi^2 B^2}{(1)} + 4\,\coef{8}{\phi^2 B^2}{(2)} & -4\,\coef{8}{\phi^2 B^2}{(3)} & \coef{8}{\phi^2 B^2}{(1)} + 4\,\coef{8}{\phi^2 B^2}{(2)} & 4\,\coef{8}{\phi^2 B^2}{(3)} \\
4\,\coef{8}{\phi^2 B^2}{(3)} & \coef{8}{\phi^2 B^2}{(1)} + 4\,\coef{8}{\phi^2 B^2}{(2)} & 4\,\coef{8}{\phi^2 B^2}{(3)} & -\coef{8}{\phi^2 B^2}{(1)} - 4\,\coef{8}{\phi^2 B^2}{(2)}
\end{pmatrix}\,.
\end{equation}
Finally, imposing $\mathbf A \succeq 0$, one obtains the constraint in Eq.~\eqref{eq:d8-pos_phi2X2}.

In the following, we present the complete set of positivity bounds associated with the dimension-8 ALP operators listed in Table~\ref{tab:dim8}.

\paragraph{$\phi^4 D^4$ class.}

The positivity bound related to this operator reads~\cite{Ananthanarayan:1994hf,Adams:2006sv}
\begin{equation}
    \coef{8}{\phi^4}{} \ge 0\,.
\end{equation}

\paragraph{$\phi^2 H^2 D^4$ class.}

\begin{figure}[h!tb]
    \centering
    \includegraphics[scale = 0.92]{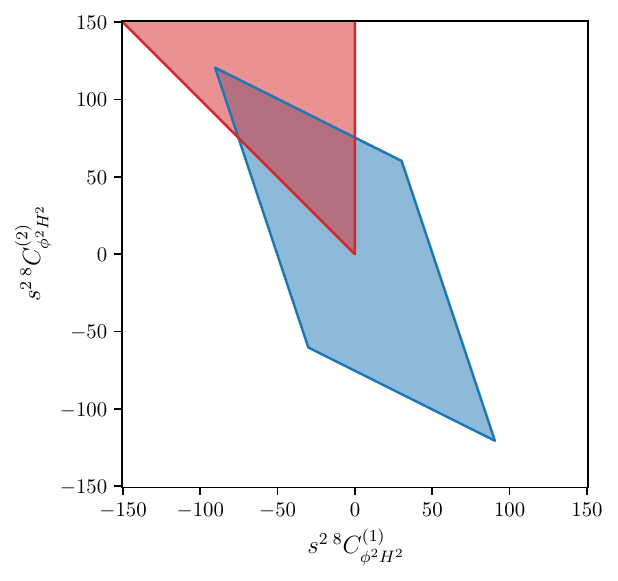}
    \caption{Parameter space allowed by partial wave unitarity bounds (Eq.~\eqref{eq:d8-uni_phi2H2}, blue) and positivity bounds (Eq.~\eqref{eq:d8-pos_phi2H2}, red) associated with the operators $\op{8}{\phi^2 H^2}{(1)}$ and $\op{8}{\phi^2 H^2}{(2)}$. The intersection area constitutes $1/4$ of the area of the blue region.}
    \label{fig:d8phi2H2}
\end{figure}

The positivity bound for this class of Wilson coefficients, 
obtained by considering all possible superpositions of scattering states, is shown in Fig.~\ref{fig:d8phi2H2} and reads
\begin{equation}
    \coef{8}{\phi^2 H^2}{(2)}-\abs{2\,\coef{8}{\phi^2 H^2}{(1)}+\coef{8}{\phi^2 H^2}{(2)}} \ge 0\,,
    \label{eq:d8-pos_phi2H2}
\end{equation}
whereas the bound obtained without including such superpositions is
\begin{equation}
    \coef{8}{\phi^2 H^2}{(2)} \ge 0\,.
\end{equation}
They are consistent with those of~\cite{Kim:2023pwf}.

\paragraph{$\phi^2 X^2 D^2$ class.}

\begin{figure}[h!tb]
    \centering
    \includegraphics[width=\linewidth]{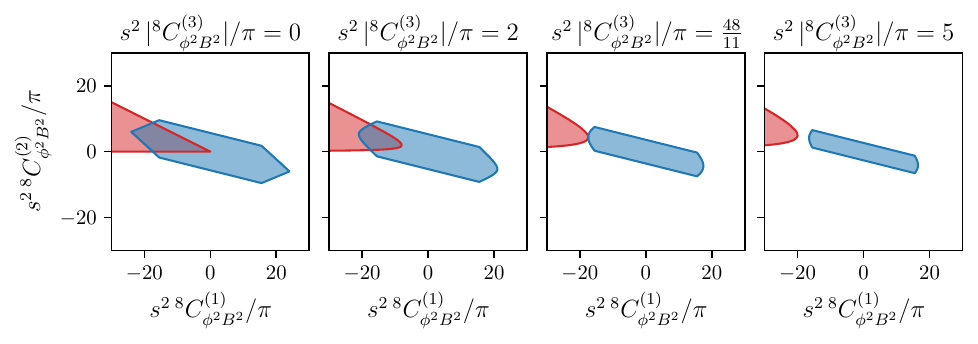}
    \\
    \includegraphics[width=\linewidth]{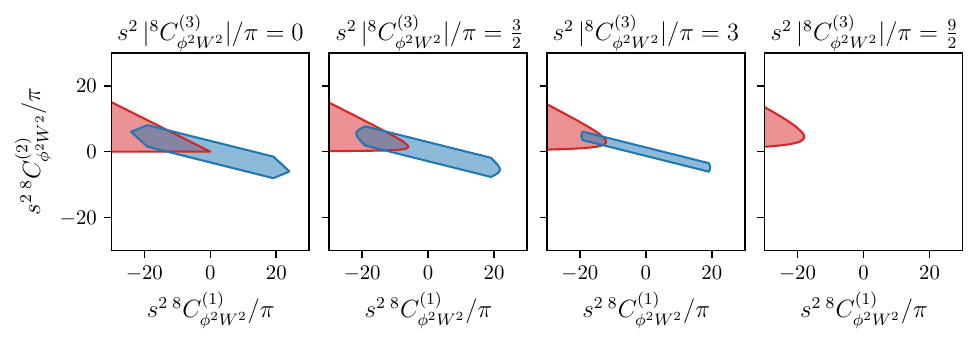}
    \\
    \includegraphics[width=\linewidth]{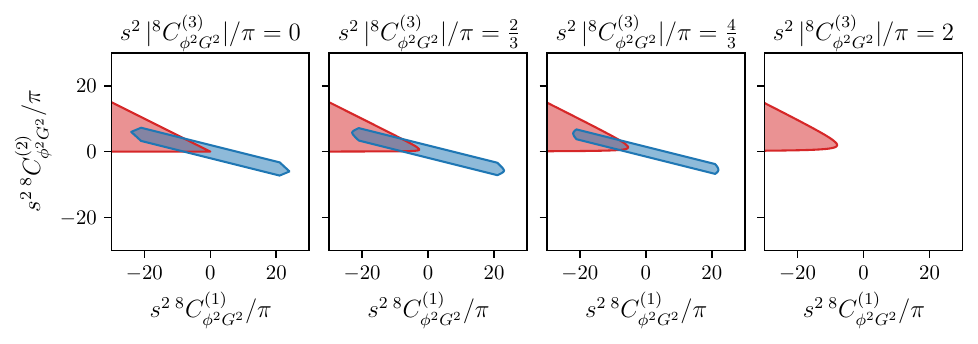}
    \caption{Parameter space allowed by the partial wave unitarity bounds 
    (Eqs.~\eqref{eq:PWUB_d8_phi2X2_1}--\eqref{eq:PWUB_d8_phi2X2_2}, blue)
    and the positivity constraints 
    (Eq.~\eqref{eq:d8-pos_phi2X2}, red)
    regarding the Wilson coefficients belonging to the class $\phi^2 X^2 D^2$ in the $\coef{8}{\phi^2 X^2}{(1)}$--$\coef{8}{\phi^2 X^2}{(2)}$ plane and computed for different values of $\abs{\coef{8}{\phi^2 X^2}{(3)}}$.}
    \label{fig:d8phi2B2}
\end{figure}

The positivity bounds for this class of Wilson coefficients, obtained by considering all possible superpositions of scattering states, are
\begin{equation}
\label{eq:d8-pos_phi2X2}
    \coef{8}{\phi^2 X^2}{(1)} + \sqrt{\left(\coef{8}{\phi^2 X^2}{(1)}+4\,\coef{8}{\phi^2 X^2}{(2)}\right)^2+\left(4\,\coef{8}{\phi^2 X^2}{(3)}\right)^2} \le 0
\end{equation}
for every $X=B,W,G$ and are shown in Fig.~\ref{fig:d8phi2B2}, whereas the bounds obtained without including such superpositions are
\begin{equation}
    \coef{8}{\phi^2 X^2}{(1)} \le 0\,.
\end{equation}

To highlight the impact of positivity bounds, we compare the volume of the parameter space allowed by unitarity (see Eqs.~\eqref{eq:PWUB_d8_phi2X2_1}--\eqref{eq:PWUB_d8_phi2X2_2}), $\text{Vol}(\mathscr U)$, with the one obtained after imposing the positivity constraints in Eq.~\eqref{eq:d8-pos_phi2X2}, $\text{Vol}(\mathscr U \cap \mathscr P)$.
In particular, for fixed values of the center-of-mass energy, we find
\begin{equation}
    \frac{\text{Vol}(\mathscr U \cap \mathscr P)}{\text{Vol}(\mathscr U)} = 
    \begin{dcases}
        \frac{36(6+\sqrt{2})}{2057}\approx 0.13 & \text{if~}d(G)=1,\\
        \frac{18\sqrt{d(G)}-11\sqrt{2}}{36\sqrt{d(G)}-6\sqrt{2}} &\text{if~}d(G)\ge 2,
    \end{dcases}
\end{equation}
depending on the dimensionality $d(G)$ of the associated gauge group $G$.

These bounds also apply if we substitute the ALP field with a Higgs doublet.
If one performs this substitution, one finds the dimension-8 SMEFT operators
\begin{align}
    Q^{(1)}_{X^2 H^2 D^2} &= (D^\mu H^\dagger D^\nu H)X^{\mathscr A}_{ \mu\rho}X_\nu^{\mathscr A\rho}\,,
    \\
    Q^{(2)}_{X^2 H^2 D^2} &= (D^\mu H^\dagger D_\mu H)X^{\mathscr A}_{ \nu\rho}X^{\mathscr A \nu \rho}\,,
    \\
    Q^{(3)}_{X^2 H^2 D^2} &= (D^\mu H^\dagger D_\mu H)X^{\mathscr A}_{ \nu\rho}\widetilde X^{\mathscr A \nu \rho}\,,
\end{align}
with $X_{\mu\nu}^{\mathscr A} = B_{\mu\nu},W_{\mu\nu}^I,G_{\mu\nu}^A$, of the basis in \cite{Murphy:2020rsh}.
It follows that the associated Wilson coefficients satisfy the constraints
\begin{equation}
    C_{X^2 H^2 D^2}^{(1)} + \sqrt{\left(C_{X^2 H^2 D^2}^{(1)}+4\,C_{X^2 H^2 D^2}^{(2)}\right)^2+\left(4\,C_{X^2 H^2 D^2}^{(3)}\right)^2} \le 0
\end{equation}
for every $X=B,W,G$,
which represent a generalization of the results presented in \cite{Chala:2023xjy,Remmen:2019cyz} that accounts for a non-vanishing CP-odd $C_{X^2 H^2 D^2}^{(3)}$.

\paragraph{$\phi^2 \psi^2 D^3$ class.}
The positivity bounds for this class of Wilson coefficients, obtained by considering all possible superpositions of scattering states, are
\begin{equation}
    \coef{8}{\phi^2\psi^2}{} \succeq 0\,,
\end{equation}
or, equivalently,
\begin{align}
\label{eq:pos_d8_phi2psi2_1}
    0\le \Tr (\coef{8}{\phi^2 \psi^2}{}) &= \coef{8}{\phi^2 \psi^2}{11}+\coef{8}{\phi^2 \psi^2}{22}+\coef{8}{\phi^2 \psi^2}{33}\,,
    \\
    0 \le \frac{1}{2}\left[\left(\Tr (\coef{8}{\phi^2 \psi^2}{}) \right)^2 - \Tr (\coef{8}{\phi^2 \psi^2}{2})\right] &= \coef{8}{\phi^2 \psi^2}{11}\,\coef{8}{\phi^2 \psi^2}{22}+\coef{8}{\phi^2 \psi^2}{11}\,\coef{8}{\phi^2 \psi^2}{33}+\coef{8}{\phi^2 \psi^2}{22}\,\coef{8}{\phi^2 \psi^2}{33}\nonumber \\&\quad -\abs{\coef{8}{\phi^2 \psi^2}{12}}^2-\abs{\coef{8}{\phi^2 \psi^2}{13}}^2-\abs{\coef{8}{\phi^2 \psi^2}{23}}^2\,, \label{eq:pos_d8_phi2psi2_2} \\
    0\le \det (\coef{8}{\phi^2 \psi^2}{}) &= \coef{8}{\phi^2 \psi^2}{11}\,\coef{8}{\phi^2 \psi^2}{22}\,\coef{8}{\phi^2 \psi^2}{33}+\coef{8}{\phi^2 \psi^2}{12}\,\coef{8}{\phi^2 \psi^2}{23}\,\coef{8}{\phi^2 \psi^2}{31}\nonumber \\
    &\quad+
    \coef{8}{\phi^2 \psi^2}{13}\,\coef{8}{\phi^2 \psi^2}{32}\,\coef{8}{\phi^2 \psi^2}{21}
    -\coef{8}{\phi^2 \psi^2}{11}\abs{\coef{8}{\phi^2 \psi^2}{23}}^2\nonumber \\
    &\quad-\coef{8}{\phi^2 \psi^2}{22}\abs{\coef{8}{\phi^2 \psi^2}{13}}^2-\coef{8}{\phi^2 \psi^2}{33}\abs{\coef{8}{\phi^2 \psi^2}{12}}^2\,, \label{eq:pos_d8_phi2psi2_3}
\end{align}
for every $\psi = \ell,e,q,u,d$.
This means that:
\begin{itemize}
    \item Every diagonal entry of $\coef{8}{\phi^2 \psi^2}{}$ must be positive;
    \item The modulus of the off-diagonal entries of $\coef{8}{\phi^2 \psi^2}{}$ is bounded from above by combinations of its diagonal entries.
\end{itemize}
Instead, if one did not consider all possible superpositions of scattering states, the resulting bound would just involve the diagonal entries:
\begin{equation}
    \coef{8}{\phi^2 \psi^2}{pp} \ge 0\,.
\end{equation}

Also in this case, by substituting the ALP field with a Higgs doublet, one recovers the dimension-8 SMEFT operators 
\begin{align}
    Q_{\psi^2 H^2 D^3}^{(1)pr} &= i(\overline \psi_p \gamma^\mu D^\nu \psi_r)(D_{(\mu}D_{\nu)} H^\dagger H)\,,
    \\
    Q_{\psi^2 H^2 D^3}^{(2)pr} &= i(\overline \psi_p \gamma^\mu D^\nu \psi_r)(H^\dagger D_{(\mu}D_{\nu)}H)\,,
\end{align}
with $\psi = \ell,e,q,u,d$, of the basis in \cite{Murphy:2020rsh}.
Identifying $\coef{8}{\phi^2\psi^2}{pr}$ with $-\frac{1}{2}C_{\psi^2 H^2 D^3}^{(1)pr}-\frac{1}{2}C_{\psi^2 H^2 D^3}^{(2)pr}$, the positivity bounds in Eqs.~\eqref{eq:pos_d8_phi2psi2_1}, \eqref{eq:pos_d8_phi2psi2_2}, and \eqref{eq:pos_d8_phi2psi2_3} are translated in
\begin{equation}\label{eq:pos_smeft_psi2H2D3}
    C^{(1)}_{\psi^2 H^2 D^3} + C^{(2)}_{\psi^2 H^2 D^3} \preceq 0\,,
\end{equation}
or, equivalently,
\begin{align}
    0 &\ge \Tr(C_{\psi^2 H^2 D^3}^{(1)}+C_{\psi^2 H^2 D^3}^{(2)})\,,\\
    0 &\le \left[\Tr(C_{\psi^2 H^2 D^3}^{(1)}+C_{\psi^2 H^2 D^3}^{(2)})\right]^2  - \Tr[\left(C_{\psi^2 H^2 D^3}^{(1)}+C_{\psi^2 H^2 D^3}^{(2)}\right)^2]\,,\\
    0 &\ge\det(C_{\psi^2 H^2 D^3}^{(1)}+C_{\psi^2 H^2 D^3}^{(2)})\,,
\end{align}
for every $\psi = \ell,e,q,u,d$,
which represent a generalization of the results presented in \cite{Chala:2023xjy} that accounts for non-vanishing off-diagonal entries of $C_{\psi^2 H^2 D^3}^{(1)}$ and $C_{\psi^2 H^2 D^3}^{(2)}$.\footnote{According to the results presented in \cite{Chala:2021wpj,Chala:2023jyx,Chala:2023xjy,Liao:2025npz}, one can notice that the positivity bound in Eq.~\eqref{eq:pos_smeft_psi2H2D3} is consistent with the fact that the one-loop renormalization group equation for the combination $C_{\psi^2 H^2 D^3}^{(1)}+C_{\psi^2 H^2 D^3}^{(2)}$, if restricted to the terms stemming from double insertions of dimension-6 operators, is positive semidefinite. In fact, from \cite{Bakshi:2024wzz}, we have that, e.g., $\dv[]{}{\ln\mu}(C_{e^2 H^2 D^3}^{(1)pr}+C_{e^2 H^2 D^3}^{(2)pr}) \supset \frac{1}{4\pi^2} C_{He}^{pq}C_{He}^{qr}$, which is positive semidefinite since $C_{He}$ is a Hermitian matrix, being the Wilson coefficient associated with the operator $Q_{He} = i(H^\dagger \overset{\leftrightarrow}{D}_\mu H)(\overline e \gamma^\mu e)$ \cite{Grzadkowski:2010es}. One can check that the same is true for all the other $\psi$.}

We note that the positivity bounds derived above are obtained at tree level. In the SMEFT context, it is known that one-loop corrections can
quantitatively modify
positivity bounds~\cite{Bellazzini:2020cot,Bellazzini:2021oaj,Chala:2021wpj,Li:2022aby,Ye:2024rzr}. In particular,
renormalization-group running of
the Wilson coefficients induces an energy-scale dependence of the
bounds, while loop
contributions from light states entering the dispersive integral can
shift the resulting
constraints. A systematic inclusion of these effects in the ALP-SMEFT
framework is beyond
the scope of the present work. Nevertheless, the tree-level bounds
derived here provide
the leading-order constraints implied by analyticity, unitarity, and
crossing symmetry.

\section{Phenomenological applications}
\label{pheno} 

In this section, we provide a few illustrative examples to emphasize the importance of imposing partial wave unitarity bounds. As we will show, the size of physical effects in several observables can be drastically reduced once we assume perturbative unitarity to hold at a given energy scale.

\subsection{Dimension-5 couplings to gauge bosons}

\begin{figure}[h!tb]
    \centering
    \includegraphics[scale = 0.92]{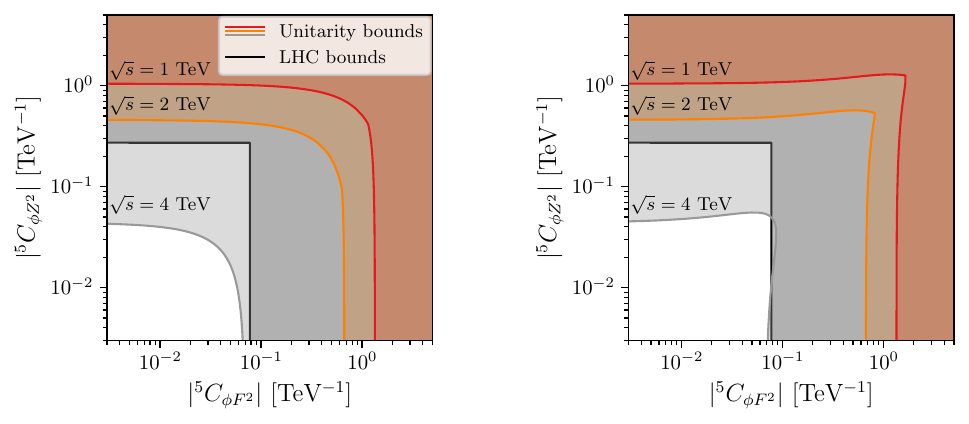}
    \caption{Parameter space excluded by the partial wave unitarity bounds in Eq.~\eqref{eq:PWUB_d5_phiX2} for the dimension-5 ALP couplings to photons ($\coef{5}{\phi F^2}{} = c^2_W \, \coef{5}{\phi B^2}{} + s^2_W \,\coef{5}{\phi W^2}{}$) and $Z$-bosons ($\coef{5}{\phi Z^2}{} = s^2_W \, \coef{5}{\phi B^2}{} + c^2_W\,\coef{5}{\phi W^2}{}$), in the cases where $\coef{5}{\phi F^2}{} \coef{5}{\phi Z^2}{} < 0$ (\textit{left panel}) and $\coef{5}{\phi F^2}{} \coef{5}{\phi Z^2}{} > 0$ (\textit{right panel}). In both cases the coupling to gluons have been fixed to $\abs{\coef{5}{\phi G^2}{}} = 0.225$~TeV$^{-1}$.
    The black region corresponds to the parameter space excluded by Ref.~\cite{Gavela:2019cmq} via non-resonant ALP searches at the LHC, where the limits $\abs{\coef{5}{\phi F^2}{}\coef{5}{\phi G^2}{}} < 0.018$~TeV$^{-2}$ and $\abs{\coef{5}{\phi Z^2}{}\coef{5}{\phi G^2}{}} < 0.061$~TeV$^{-2}$ are found for $m_\phi \lesssim 200$~GeV.}
    \label{fig:gaugebosons}
\end{figure}

As a first example, we consider the partial wave unitarity bounds 
for the dimension-5 ALP couplings to gauge bosons. As discussed in Ref.~\cite{Gavela:2019cmq}, non-resonant ALP searches at the LHC
are particularly effective in this case, leading to the excluded 
black regions of Fig.~\ref{fig:gaugebosons}.
Depending on the assumed center-of-mass energy $\sqrt{s}$, unitarity bounds can be competitive or superior to the experimental bounds to probe the ALP parameter space. 

\subsection{\texorpdfstring{Dimension-7 couplings in $l^+ l^- \to \phi h$}{Dimension-7 couplings in l+ l- -> phi h}}

\begin{figure}[h!tb]
    \centering
    \includegraphics[scale = 0.92]{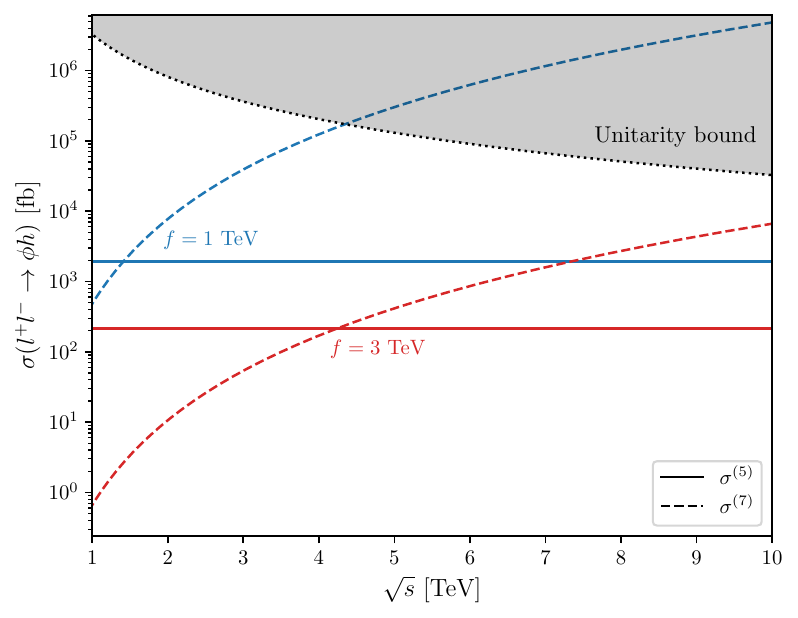}
    \caption{Cross section for the process $l^+ l^- \to \phi h$, with $l=e,\mu$, in the high-energy limit. The relevant Wilson coefficients are parameterized as $\coef{5}{\phi \ell e H}{ll} = i \, y_l/f$ and $\coef{7}{\phi \ell e H}{(1)ll} = \coef{7}{\phi \ell e H}{(2)ll} = 1/f^3$, where $f$ can be interpreted as the ALP decay constant.}
    \label{fig:crosssection}
\end{figure}

The relevance of higher-dimensional ALP operators can be nicely illustrated through the process $l^+ l^- \to \phi h$~\cite{Grojean:2023tsd}, which receives contributions from both dimension-5 and -7 operators.
In Fig.~\ref{fig:crosssection}, we show the cross section for this process in the high-energy limit. The relevant Wilson coefficients are parameterized as $\coef{5}{\phi \ell e H}{ll} = i \, y_l/f$ and $\coef{7}{\phi \ell e H}{(1)ll} = \coef{7}{\phi \ell e H}{(2)ll} = 1/f^3$, where $f$ can be interpreted as the ALP decay constant.

In the high-energy limit, the cross section of $l^+ l^- \to \phi h$ 
is $\sigma = \sigma^{(5)} + \sigma^{(7)}$, where
\begin{equation}
    \sigma^{(7)} = \frac{s^2}{768\pi}\left[\abs{\coef{7}{\phi \ell e H}{(1)ll}}^2+\abs{\coef{7}{\phi \ell e H}{(2)ll}}^2+\Re\left(\coef{7}{\phi \ell e H}{(1)ll}\,\coef{7}{\phi \ell e H}{(2)ll}{}^*\right)\right],
\quad
    \sigma^{(5)} = \frac{1}{64\pi}\abs{\coef{5}{\phi \ell e H}{ll}}^2\,.
\end{equation}
On the other hand, the partial wave unitarity bounds lead to the following conditions 
\begin{equation}
    \sigma^{(7)} \le \frac{5\pi}{3s}\,,\qquad\qquad
    \sigma^{(5)} \le \frac{\pi}{s}\,,
\end{equation}
that are illustrated in Fig.~\ref{fig:crosssection}.
As a result, large effects to the $l^+ l^- \to \phi h$ cross section can be significantly reduced depending on the ALP decay constant and the center-of-mass energy.

\subsection{Weak-violating ALPs and rare decays}

In the case of weak-violating ALP interacting with electrons (with $\overline g_{ee} = g_{\nu_e} = 0$), the partial wave unitarity bound reads
\begin{equation}
    \abs{g_{ee}} \lesssim 10^{-5} \left(\frac{\text{TeV}}{\sqrt{s}}\right)^2\,.
\end{equation}
Instead, in the weak-preserving case (with $\overline g_{ee} = g_{ee}$ and $g_{\nu_e} = 0$), we have
\begin{equation}
    \abs{g_{ee}} \lesssim 4 \left(\frac{\text{TeV}}{\sqrt{s}}\right)\,.
\end{equation}

In order to monitor the impact of the above bounds on physical observables, we consider the branching fractions of $\pi^+\to e^+\nu_e \phi$, $K^+\to e^+\nu_e \phi$, and $W^+\to e^+\nu_e \phi$ decays~\cite{Altmannshofer:2022ckw} in the weak-violating case. 
Neglecting ALP mass effects and taking into account partial wave unitarity bounds, we find that
\begin{align}
    \mathcal B(\pi^+ \to e^+ \nu_e \phi) &\lesssim 4 \times 10^{-9}\left(\frac{\text{TeV}}{\sqrt{s}}\right)^4 \,,
    \\
    \mathcal B(K^+ \to e^+ \nu_e \phi) &\lesssim 7 \times 10^{-8} \left(\frac{\text{TeV}}{\sqrt{s}}\right)^4\,,
    \\
    \mathcal B(W^+ \to e^+ \nu_e \phi) &\lesssim 6 \times 10^{-5} \left(\frac{\text{TeV}}{\sqrt{s}}\right)^4\,.
\end{align}
The current bound for the 
$\pi^+ \to e^+ \nu_e \phi$ process arises from the SINDRUM experiment which looked for $e^+e^-$ resonances in $\pi^+ \to e^+ \nu_e \phi$ followed by 
$\phi \to e^+e^-$
with sensitivity to branching ratios of $\mathcal{O}(10^{-10})$~\cite{SINDRUM:1986klz}.
In the future, the PIONEER experiment aims to reach 
the sensitivity of $\mathcal{O}(10^{-11})$~\cite{PIONEER:2022yag}.
Concerning the $K^+ \to e^+ \nu_e \phi$ decay, 
we exploit the E865 search for the SM process 
$K^+ \to e^+ \nu_e e^+e^-$~\cite{Poblaguev:2002ug}. 
Assuming that the ALP channel does not exceed 
twice the uncertainty in this measurement, we impose
the bound $\mathcal{B}(K^+ \to e^+ \nu_e \phi) \leq 4 \times 10^{-9}$~\cite{Altmannshofer:2022ckw}. 
As a future projection, we assume a reference 
bound of $10^{-10}$~\cite{Altmannshofer:2022ckw}.
Moreover, following the analysis of Ref.~\cite{Altmannshofer:2022ckw}, we also impose 
the constraints from leptonic charged meson decays in the CHARM experiment~\cite{CHARM:1985anb}.
As for the $W^+ \to e^+ \nu_e \phi$ decay, we require this mode not to contribute at a visible level to the total width of the $W$ boson
$\Gamma_W = 2.085 \pm 0.042$~GeV~\cite{ParticleDataGroup:2022pth}.
Dedicated analyses for other rare $W$ decays by CMS could improve the bound on the branching ratios of $W^+ \to e^+ \nu_e \phi$ at the level of $10^{-6}$~\cite{Altmannshofer:2022ckw}.

\begin{figure}[h!tb]
    \centering
    \includegraphics[scale = 0.92]{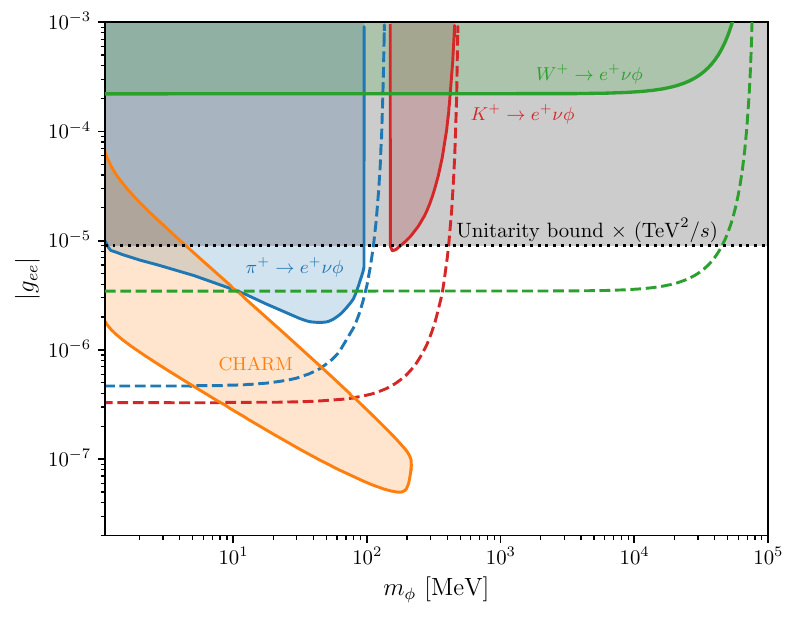}
    \caption{Theoretical and experimental bounds on the coupling $g_{ee}$ of a weak-violating ALP~\cite{Altmannshofer:2022ckw}. 
    The partial wave unitarity bound is shown in grey and scales 
    as $1/s$. 
    Searches for leptonic charged meson decays are shown in blue (pions)~\cite{PIONEER:2022yag} and red (kaons)~\cite{Poblaguev:2002ug,Altmannshofer:2022ckw}, while searches for rare $W$ boson decays are shown in green~\cite{ParticleDataGroup:2022pth,Altmannshofer:2022ckw}. Bounds from leptonic charged meson decays in the CHARM proton beam dump~\cite{CHARM:1985anb} are shown in orange~\cite{Altmannshofer:2022ckw}. Dashed lines refer to the potential sensitivities with dedicated searches at the PIONEER experiment~\cite{PIONEER:2022yag} and kaon factories~\cite{Altmannshofer:2022ckw}.
    }
    \label{fig:WV_gee}
\end{figure}

As shown in Fig.~\ref{fig:WV_gee}, unitarity bounds already exclude large regions of the parameter space for $\sqrt{s} = 1$~TeV and are typically stronger than the current experimental bounds on charged meson as well as $W$ boson decays.

\section{Conclusions}
\label{conclusions}
Axion-like particles (ALPs) are generic pseudo-Nambu-Goldstone bosons (pNGBs) emerging from the spontaneous breaking of some global symmetry 
above the electroweak scale. The pNGB nature of ALPs forces their interactions with Standard Model fields to be derivative. Therefore, ALP interactions exhibit an inherent growth with the energy, possibly leading to the breakdown of perturbative unitarity at some large energy scale. 

In this work, we have systematically evaluated partial wave unitarity bounds for ALP Effective Field Theories (EFTs) including higher-dimensional operators up to dimension 8. The adopted methodology is based on a recently developed formalism, based on on-shell methods, which has proven particularly efficient for studying unitarity bounds of EFTs at high energies~\cite{Bresciani:2025toe}.

In particular, we first revisited and improved the bounds of Ref.~\cite{Brivio:2021fog} relative to dimension-5 and -6 operators, finding that the processes $XX \to XX$ and $XX \to \phi \phi$ (with $X=B,W,G$) set the most stringent bounds at any energy scale.
The summary of the partial wave unitarity bounds on the individual
Wilson coefficients associated with these operators is reported in 
Fig.~\ref{fig:d56summary}.
Stronger bounds can be obtained through a coupled-channel analysis, 
where several scattering processes such as $BB \to WW$, $WW \to GG$, 
and $BB \to GG$ are generated by the simultaneous presence of different
Wilson coefficients; see Fig.~\ref{fig:d5phiX2}.
As a relevant phenomenological application, we discussed the 
impact of the above bounds on non-resonant ALP searches at the LHC~\cite{Gavela:2019cmq}. Unitarity bounds can be competitive or even more stringent than the experimental bounds already for center-of-mass energies at the TeV scale; see Fig.~\ref{fig:gaugebosons}. 

Moreover, we discussed partial wave unitarity bounds for dimension-5 
weak-violating ALP interactions, see Fig.~\ref{fig:WV}, as in this 
case there is a striking energy enhancement $({\rm energy}/m_l)$ 
in several processes such as charged meson and $W$ boson decays~\cite{Altmannshofer:2022ckw}.
As shown in Fig.~\ref{fig:WV_gee}, unitarity bounds exclude large regions of the parameter space already for $\sqrt{s} = 1$ TeV and 
are typically stronger than the current experimental bounds.

Finally, we analyzed the partial wave unitarity bounds for dimension-7 and -8 operators which were recently classified in~\cite{Song:2023lxf,Grojean:2023tsd,Bertuzzo:2023slg}; see
Figs.~\ref{fig:d7phipsi2H}, \ref{fig:d7summary}, and \ref{fig:d8summary}.
As a phenomenological application, we considered the process $l^+ l^- \to \phi h$, which receives contributions from both dimension-5 and -7 operators, and which can be looked for at future high-energy lepton colliders~\cite{Grojean:2023tsd}. As shown in Fig.~\ref{fig:crosssection}, potentially large effects by dimension-7 operators are severely reduced depending on the ALP decay constant and the center-of-mass energy.

The last part of this work was devoted to the derivation of positivity bounds, stemming from the analyticity and causality of scattering amplitudes, for dimension-8 operators. These were derived from elastic $2 \to 2$ scattering processes that grow with energy as $s^2$.
As shown in Figs.~\ref{fig:d8phi2H2} and \ref{fig:d8phi2B2}, positivity and partial wave unitarity bounds are highly complementary in constraining the parameter space of ALP EFTs. 
Interestingly, our results can also be used to derive new positivity constraints in the SMEFT; see Section~\ref{sec:positivity}.

To conclude, the positivity and partial wave unitarity bounds derived in our study can be applied in several directions, such as ALP searches at colliders or in a variety of rare decays
or even as constraints for the event generation which can impact the shapes of the expected distributions of experimental searches. 
Our work highlights the synergy and interplay of first-principle-based theoretical bounds and experimental limits in our adventure to discover new physics phenomena.

\acknowledgments
This work received funding by the INFN Iniziative Specifiche AMPLITUDES and APINE and from the European Union’s Horizon 2020 research and innovation programme under the Marie Sklodowska-Curie grant agreements n. 860881 – HIDDeN, n. 101086085 – ASYMMETRY. This work was also partially supported by the Italian MUR Departments of Excellence grant 2023-2027 “Quantum Frontiers”. The work of P.P. is supported by the European Union – Next Generation EU and by the Italian Ministry of University and Research (MUR) via the PRIN 2022 project n. 2022K4B58X – AxionOrigins. G.L. gratefully acknowledges financial support from the Swiss National Science Foundation (Project No. TMCG-2 213690).

\appendix

\section{Conventions}
\label{sec:conventions}

Concerning the convention for the Minkowski metric, which is relevant for setting and interpreting the positivity bounds, we adopt the mostly-minus signature.

As for the helicity spinors $\lambda_{\alpha}$ and $\widetilde\lambda^{\dot\alpha}$, they transform in the $(1/2,0)$ and $(0,1/2)$ representations of $SL(2,\mathbb{C})$, respectively.
Spinor indices are raised and lowered according to $\lambda^\alpha = \epsilon^{\alpha\beta}\lambda_\beta$ and $\widetilde \lambda_{\dot\alpha} = \epsilon_{\dot \alpha \dot \beta} \widetilde\lambda^{\dot\beta}$, where we use the convention $\epsilon^{12} = \epsilon^{\dot 1 \dot 2} = -\epsilon_{12} =-\epsilon_{\dot 1 \dot 2} = 1$.
The spinor decomposition of a light-like four-momentum $p_{\mu}$ of an outgoing particle is given by
\begin{equation}
    p_{\alpha \dot\alpha} = p_\mu \sigma^{\mu}_{\alpha \dot\alpha} = \lambda_\alpha \widetilde \lambda_{\dot \alpha}\,,
\end{equation}
where $\sigma^\mu = (1,\vec \sigma)$ and $\sigma^I$ are the Pauli matrices.
The Lorentz invariant angle and square inner products are then defined as
\begin{equation}
    \agl{i}{j} = \lambda^\alpha_i \lambda_{j\,\alpha} = \epsilon_{\alpha\beta}\lambda_i^\alpha \lambda_j^\beta\,,
    \qquad
    \sqr{i}{j} = \widetilde \lambda_{i\,\dot\alpha}\widetilde\lambda_j^{\dot\alpha} = -\epsilon_{\dot\alpha \dot\beta}\widetilde\lambda_{i}^{\dot\alpha} \widetilde\lambda_j^{\dot\beta}\,,
\end{equation}
in such a way that Mandelstam invariants can be written as $s_{ij} = 2 p_i \cdot p_j = \agl{i}{j}\sqr{j}{i}$.
Within this formalism, polarization vectors can be represented as
\begin{align}
    \varepsilon^\mu_-(p) = \frac{\langle p\,\sigma^\mu \, q]}{\sqrt{2}\sqr{p}{q}}\,, \qquad
    \varepsilon^\mu_+(p) = \frac{\langle q\,\sigma^\mu \, p]}{\sqrt{2}\agl{q}{p}}\,,
\end{align}
where $q$ is a reference momentum such that $\sqr{p}{q}, \agl{q}{p} \neq 0$, while spinors associated with Dirac fermions are
\begin{align}
    u_+(p) &= v_-(p) = 
    \begin{pmatrix}
        \lambda_\alpha \\ 0
    \end{pmatrix}
    \,,
    &
    u_-(p) &= v_+(p) = 
    \begin{pmatrix}
        0 \\ \widetilde\lambda^{\dot\alpha}
    \end{pmatrix}
    \,, 
    \\
    \overline u_+(p) &= \overline v_-(p) = 
    \begin{pmatrix}
        0 & \widetilde\lambda_{\dot\alpha}
    \end{pmatrix}
    \,,
    &
    \overline u_-(p) &= \overline v_+(p) = 
    \begin{pmatrix}
        \lambda^\alpha & 0
    \end{pmatrix}
    \,.
\end{align}

\section{Amplitudes}
\label{Appendix_A}

In this appendix, we list the relevant amplitudes considered for obtaining the most stringent partial wave unitarity bounds reported in the main text.

All particles are considered outgoing and massless, since we are interested in the high-energy limit, and the superscripts $\pm$ denote the signs of helicities. 
The following notation is used: 
\begin{itemize}
    \item $p,r,\dotsc\in\{1,2,3\}$ are weak-eigenstate indices;
    \item $i,j,\dotsc\in\{1,2\}$ are $SU(2)$ fundamental indices;
    \item $I,J,\dotsc \in \{1,2,3\}$ are $SU(2)$ adjoint indices;
    \item $\alpha,\beta,\dotsc \in\{1,2,3\}$ are $SU(3)$ fundamental indices;
    \item $A,B,\dotsc \in \{1,\dotsc,8\}$ are $SU(3)$ adjoint indices.
\end{itemize}

\subsection{Dimension-5 and -6 interactions\label{app:amplitudes5e6}}

The relevant amplitudes generated by one or two insertions of dimension-5 operators (see Table~\ref{tab:dim5-6}) are 
\begin{align}
\mathcal A(\phi,\ell^-_{ip},\overline e^-_{r},\overline H_j) &= \coef{5}{\phi \ell e H}{pr} \,\agl{2}{3}\delta_{ij} \,, 
\\
\mathcal A(\phi,q^-_{\alpha ip},\overline d^-_{\beta r},\overline H_j) &= \coef{5}{\phi qd H}{pr} \,\agl{2}{3}\delta_{\alpha\beta}\delta_{ij}\,, 
\\
\mathcal A(\phi,q^-_{\alpha ip},\overline u^-_{\beta r}, H_j) &= \coef{5}{\phi qu H}{pr} \,\agl{2}{3}\delta_{\alpha\beta}\epsilon_{ij}\,, 
\\
    \mathcal A(B^+,B^+,B^-,B^-) &= - 4\, \coef{5}{\phi B^2}{2}\,\frac{\agl{3}{4}^2 \sqr{2}{1}}{\agl{1}{2}}\,,  
    \\
    \mathcal A(\phi,\phi,B^-,B^-) &= 8\,\coef{5}{\phi B^2}{2}\,\agl{3}{4}^2\,,
    \\
    \mathcal A(W_I^+,W_J^+,W_K^-,W_L^-) &= - 4\, \coef{5}{\phi W^2}{2}\,\frac{\agl{3}{4}^2 \sqr{2}{1}}{\agl{1}{2}}\delta^{IJ}\delta^{KL}\,, 
    \\
    \mathcal A(W_I^+,W_J^+,W_K^+,W_L^+) &= 4\, \coef{5}{\phi W^2}{2} \bigg[\frac{\sqr{2}{1}\sqr{4}{3}^2}{\agl{1}{2}}\delta^{IJ}\delta^{KL}+\frac{\sqr{3}{1}\sqr{4}{2}^2}{\agl{1}{3}}\delta^{IK}\delta^{JL}\nonumber \\&\quad +\frac{\sqr{4}{1}\sqr{3}{2}^2}{\agl{1}{4}}\delta^{IL}\delta^{JK}\bigg]\,, 
    \\
    \mathcal A(\phi,\phi,W_I^-,W_J^-) &= 8\,\coef{5}{\phi W^2}{2}\,\agl{3}{4}^2 \delta^{IJ}\,,
    \\
    \mathcal A(G_A^+,G_B^+,G_C^-,G_D^-) &= - 4\, \coef{5}{\phi G^2}{2}\,\frac{\agl{3}{4}^2 \sqr{2}{1}}{\agl{1}{2}}\delta^{AB}\delta^{CD}\,, 
    \\
    \mathcal A(G_A^+,G_B^+,G_C^+,G_D^+) &= 4\, \coef{5}{\phi G^2}{2} \bigg[\frac{\sqr{2}{1}\sqr{4}{3}^2}{\agl{1}{2}}\delta^{AB}\delta^{CD}+\frac{\sqr{3}{1}\sqr{4}{2}^2}{\agl{1}{3}}\delta^{AC}\delta^{BD}\nonumber \\&\quad+\frac{\sqr{4}{1}\sqr{3}{2}^2}{\agl{1}{4}}\delta^{AD}\delta^{BC}\bigg] \,, 
    \\
    \mathcal A(\phi,\phi,G_A^-,G_B^-) &= 8\,\coef{5}{\phi G^2}{2}\,\agl{3}{4}^2 \delta^{AB}\,,
    \\
    \mathcal A(B^+,B^+,W_I^-,W_J^-) &= - 4\, \coef{5}{\phi B^2}{}\coef{5}{\phi W^2}{}\,\frac{\agl{3}{4}^2 \sqr{2}{1}}{\agl{1}{2}}\delta^{IJ} \,, 
    \\
    \mathcal A(B^+,B^+,W_I^+,W_J^+) &=  4\, \coef{5}{\phi B^2}{}\coef{5}{\phi W^2}{}\,\frac{\sqr{4}{3}^2 \sqr{2}{1}}{\agl{1}{2}}\delta^{IJ} \,, 
    \\
    \mathcal A(B^+,B^+,G_A^-,G_B^-) &= - 4\, \coef{5}{\phi B^2}{}\coef{5}{\phi G^2}{}\,\frac{\agl{3}{4}^2 \sqr{2}{1}}{\agl{1}{2}}\delta^{AB} \,, 
    \\
    \mathcal A(B^+,B^+,G_A^+,G_B^+) &=  4\, \coef{5}{\phi B^2}{}\coef{5}{\phi G^2}{}\,\frac{\sqr{4}{3}^2 \sqr{2}{1}}{\agl{1}{2}}\delta^{AB} \,, 
    \\
    \mathcal A(W_I^+,W_J^+,G_A^-,G_B^-) &= - 4\, \coef{5}{\phi W^2}{}\coef{5}{\phi G^2}{}\,\frac{\agl{3}{4}^2 \sqr{2}{1}}{\agl{1}{2}}\delta^{IJ}\delta^{AB} \,, 
    \\
    \mathcal A(W_I^+,W_J^+,G_A^+,G_B^+) &=  4\, \coef{5}{\phi W^2}{}\coef{5}{\phi G^2}{}\,\frac{\sqr{4}{3}^2 \sqr{2}{1}}{\agl{1}{2}}\delta^{IJ}\delta^{AB}\,.
\end{align}

The relevant amplitude generated by the only dimension-6 operator (see Table~\ref{tab:dim5-6}) is 
\begin{equation}
    \mathcal A(\phi,\phi,H_i,\overline H_j) = -\coef{6}{\phi^2 H^2}{} \,\agl{1}{2}\sqr{2}{1}\delta_{ij}\,.
\end{equation}

\subsection{Dimension-7 interactions}

The relevant amplitudes generated by one insertion of dimension-7 operators (see Table~\ref{tab:dim7}) are 
\begin{align}
    \mathcal A(\phi,\ell^-_{ip},\overline e_r^-,\overline H_j)&=-\frac{1}{2}\left(\coef{7}{\phi \ell e H}{(1)pr} \, \agl{1}{3}\sqr{3}{1} + \coef{7}{\phi \ell e H}{(2)pr} \, \agl{1}{2}\sqr{2}{1}\right)\agl{2}{3}\delta_{ij}\,,
    \\
    \mathcal A(\phi,q^-_{\alpha ip},\overline d_{\beta r}^-,\overline H_j)&=-\frac{1}{2}\left(\coef{7}{\phi qd H}{(1)pr} \, \agl{1}{3}\sqr{3}{1} + \coef{7}{\phi qd H}{(2)pr} \, \agl{1}{2}\sqr{2}{1}\right)\agl{2}{3}\delta_{\alpha\beta}\delta_{ij}\,,
    \\
    \mathcal A(\phi,q^-_{\alpha ip},\overline u_{\beta r}^-, H_j)&=-\frac{1}{2}\left(\coef{7}{\phi qu H}{(1)pr} \, \agl{1}{3}\sqr{3}{1} + \coef{7}{\phi qu H}{(2)pr} \, \agl{1}{2}\sqr{2}{1}\right)\agl{2}{3}\delta_{\alpha\beta}\epsilon_{ij}\,,
    \\
    \mathcal A(\phi,\ell_{ip}^-,\overline \ell_{jr}^+,B^-) &= \frac{1}{\sqrt{2}}\left(\coef{7}{\phi \ell^2 B}{(1)pr} + i\,\coef{7}{\phi \ell^2 B}{(2)pr}\right) \agl{1}{4}\agl{2}{4}\sqr{3}{1}\delta_{ij} \,,
    \\
    \mathcal A(\phi,e_{p}^+,\overline e_{r}^-,B^-) &= \frac{1}{\sqrt{2}}\left(\coef{7}{\phi e^2 B}{(1)pr} + i\,\coef{7}{\phi e^2 B}{(2)pr}\right) \agl{1}{4}\agl{3}{4}\sqr{2}{1} \,,
    \\
    \mathcal A(\phi,q_{\alpha ip}^-,\overline q_{\beta jr}^+,B^-) &= \frac{1}{\sqrt{2}}\left(\coef{7}{\phi q^2 B}{(1)pr} + i\,\coef{7}{\phi q^2 B}{(2)pr}\right) \agl{1}{4}\agl{2}{4}\sqr{3}{1}\delta_{\alpha\beta}\delta_{ij} \,,
    \\
    \mathcal A(\phi,u_{\alpha p}^+,\overline u_{\beta r}^-,B^-) &= \frac{1}{\sqrt{2}}\left(\coef{7}{\phi u^2 B}{(1)pr} + i\,\coef{7}{\phi u^2 B}{(2)pr}\right) \agl{1}{4}\agl{3}{4}\sqr{2}{1}\delta_{\alpha\beta} \,,
    \\
    \mathcal A(\phi,d_{\alpha p}^+,\overline d_{\beta r}^-,B^-) &= \frac{1}{\sqrt{2}}\left(\coef{7}{\phi d^2 B}{(1)pr} + i\,\coef{7}{\phi d^2 B}{(2)pr}\right) \agl{1}{4}\agl{3}{4}\sqr{2}{1}\delta_{\alpha\beta} \,,
    \\
    \mathcal A(\phi,\ell_{ip}^-,\overline \ell_{jr}^+,W_I^-) &= \frac{1}{\sqrt{2}}\left(\coef{7}{\phi \ell^2 W}{(1)pr} + i\,\coef{7}{\phi \ell^2 W}{(2)pr}\right) \agl{1}{4}\agl{2}{4}\sqr{3}{1}\sigma^I_{ij} \,,
    \\
    \mathcal A(\phi,q_{\alpha ip}^-,\overline q_{\beta jr}^+,W_I^-) &= \frac{1}{\sqrt{2}}\left(\coef{7}{\phi q^2 W}{(1)pr} + i\,\coef{7}{\phi q^2 W}{(2)pr}\right) \agl{1}{4}\agl{2}{4}\sqr{3}{1}\delta_{\alpha\beta}\sigma^I_{ij} \,,
    \\
    \mathcal A(\phi,q_{\alpha ip}^-,\overline q_{\beta jr}^+,G_A^-) &= \frac{1}{\sqrt{2}}\left(\coef{7}{\phi q^2 G}{(1)pr} + i\,\coef{7}{\phi q^2 G}{(2)pr}\right) \agl{1}{4}\agl{2}{4}\sqr{3}{1}\lambda^A_{\alpha\beta}\delta_{ij} \,,
    \\
    \mathcal A(\phi,u_{\alpha p}^+,\overline u_{\beta r}^-,G_A^-) &= \frac{1}{\sqrt{2}}\left(\coef{7}{\phi u^2 G}{(1)pr} + i\,\coef{7}{\phi u^2 G}{(2)pr}\right) \agl{1}{4}\agl{3}{4}\sqr{2}{1}\lambda^A_{\alpha\beta} \,,
    \\
    \mathcal A(\phi,d_{\alpha p}^+,\overline d_{\beta r}^-,G_A^-) &= \frac{1}{\sqrt{2}}\left(\coef{7}{\phi d^2 G}{(1)pr} + i\,\coef{7}{\phi d^2 G}{(2)pr}\right) \agl{1}{4}\agl{3}{4}\sqr{2}{1}\lambda^A_{\alpha\beta} \,,
    \\
    \mathcal A(\phi,H_i,\overline H_j,B^-) &= \frac{1}{2\sqrt{2}}\left(\coef{7}{\phi B H^2}{(1)} + i \,\coef{7}{\phi B H^2}{(2)}\right)\agl{1}{4}(\agl{2}{4}\sqr{2}{1}-\agl{3}{4}\sqr{3}{1})\delta_{ij}\,,
    \\
    \mathcal A(\phi,H_i,\overline H_j,W_I^-) &= \frac{1}{2\sqrt{2}}\left(\coef{7}{\phi W H^2}{(1)} + i \,\coef{7}{\phi W H^2}{(2)}\right)\agl{1}{4}(\agl{2}{4}\sqr{2}{1}-\agl{3}{4}\sqr{3}{1})\sigma^I_{ij}\,,
    \\
    \mathcal A(\phi,H_i,\overline H_j,H_k,\overline H_l) &= \frac{i}{2}\,\coef{7}{\phi H^4}{}\left(-\agl{1}{2}\sqr{2}{1}+\agl{1}{3}\sqr{3}{1}-\agl{1}{4}\sqr{4}{1}+\agl{1}{5}\sqr{5}{1}\right)\nonumber \\
    &\quad \times (\delta_{ij}\delta_{kl}+\delta_{il}\delta_{kj})\,,
    \\
    \mathcal A(\phi, e^+_p,\overline e^-_r, H_i,\overline H_j) &= i\,\coef{7}{\phi e^2 H^2}{pr}\,\agl{1}{3}\sqr{2}{1}\delta_{ij}\,,
    \\
    \mathcal A(\phi, u^+_{\alpha p},\overline u^-_{\beta r}, H_i,\overline H_j) &= i\,\coef{7}{\phi u^2 H^2}{pr}\,\agl{1}{3}\sqr{2}{1}\delta_{\alpha\beta}\delta_{ij}\,,
    \\
    \mathcal A(\phi, d^+_{\alpha p},\overline d^-_{\beta r}, H_i,\overline H_j) &= i\,\coef{7}{\phi d^2 H^2}{pr}\,\agl{1}{3}\sqr{2}{1}\delta_{\alpha\beta}\delta_{ij}\,,
    \\
    \mathcal A (\phi,\ell^-_{ip},\overline \ell^+_{jr},H_k,\overline H_l) &= i\left[\left(\coef{7}{\phi \ell^2 H^2}{(1)pr}-\coef{7}{\phi \ell^2 H^2}{(2)pr}\right)\delta_{ij}\delta_{kl}+2\,\coef{7}{\phi \ell^2 H^2}{(2)pr}\,\delta_{il}\delta_{kj}\right]\agl{1}{2}\sqr{3}{1}\,,
    \\
    \mathcal A (\phi,q^-_{\alpha ip},\overline q^+_{\beta jr},H_k,\overline H_l) &= i\left[\left(\coef{7}{\phi q^2 H^2}{(1)pr}-\coef{7}{\phi q^2 H^2}{(2)pr}\right)\delta_{ij}\delta_{kl}+2\,\coef{7}{\phi q^2 H^2}{(2)pr}\,\delta_{il}\delta_{kj}\right]\agl{1}{2}\sqr{3}{1}\delta_{\alpha\beta}\,.
\end{align}

\subsection{Dimension-8 interactions}

The relevant amplitudes generated by one insertion of dimension-8 operators (see Table~\ref{tab:dim8}) are 
\begin{align}
    \mathcal A(\phi,\phi,\phi,\phi) &= 2\,\coef{8}{\phi^4}{}\,(\agl{1}{2}^2\sqr{2}{1}^2 + \agl{1}{3}^2\sqr{3}{1}^2 + \agl{1}{4}^2\sqr{4}{1}^2)\,,
    \\
    \mathcal A(\phi,\phi,H_i,\overline H_j,H_k,\overline H_l) &= -2\,\coef{8}{\phi^2 H^4}{}\,\agl{1}{2}\sqr{2}{1}(\delta_{ij}\delta_{kl}+\delta_{il}\delta_{kj})\,,
    \\
    \mathcal A(\phi,\phi,H_i,\overline H_j) &= \frac{1}{2}\,\coef{8}{\phi^2 H^2}{(1)}\,\agl{1}{2}^2\sqr{2}{1}^2 + \frac{1}{4}\,\coef{8}{\phi^2 H^2}{(2)}\,(\agl{1}{3}^2\sqr{3}{1}^2 + \agl{1}{4}^2\sqr{4}{1}^2)\,,
    \\
    \mathcal A(\phi,\phi,B^-,B^-) &= 2\left(\frac{1}{4}\,\coef{8}{\phi^2 B^2}{(1)}+\coef{8}{\phi^2 B^2}{(2)} + i\,\coef{8}{\phi^2 B^2}{(3)}\right)\agl{3}{4}^3 \sqr{4}{3}\,,
    \label{eq:ampl_d8_phi2B2(1)}
    \\
    \mathcal A(\phi,\phi,B^-,B^+) &= \coef{8}{\phi^2 B^2}{(1)}\,\agl{1}{3}^2\sqr{4}{1}^2\,,
    \label{eq:ampl_d8_phi2B2(2)}
    \\
    \mathcal A(\phi,\phi,W_I^-,W_J^-) &= 2\left(\frac{1}{4}\,\coef{8}{\phi^2 W^2}{(1)}+\coef{8}{\phi^2 W^2}{(2)} + i\,\coef{8}{\phi^2 W^2}{(3)}\right)\agl{3}{4}^3 \sqr{4}{3}\delta^{IJ}\,,
    \\
    \mathcal A(\phi,\phi,W_I^-,W_J^+) &= \coef{8}{\phi^2 W^2}{(1)}\,\agl{1}{3}^2\sqr{4}{1}^2\delta^{IJ} \,,
    \\
    \mathcal A(\phi,\phi,G_A^-,G_B^-) &= 2\left(\frac{1}{4}\,\coef{8}{\phi^2 G^2}{(1)}+\coef{8}{\phi^2 G^2}{(2)} + i\,\coef{8}{\phi^2 G^2}{(3)}\right)\agl{3}{4}^3 \sqr{4}{3}\delta^{AB}\,,
    \\
    \mathcal A(\phi,\phi,G_A^-,G_B^+) &= \coef{8}{\phi^2 G^2}{(1)}\,\agl{1}{3}^2\sqr{4}{1}^2 \delta^{AB} \,,
    \\
    \mathcal A(\phi,\phi,\ell^-_{ip},\overline \ell^+_{jr}) &= \frac{1}{2}\,\coef{8}{\phi^2 \ell^2}{pr}\left(\agl{1}{3}\agl{1}{4}\sqr{4}{1}^2 + \agl{2}{3}\agl{2}{4}\sqr{4}{2}^2\right)\delta_{ij}\,,
    \\
    \mathcal A(\phi,\phi,e^+_p,\overline e^-_r) &= \frac{1}{2}\,\coef{8}{\phi^2 e^2}{pr}\left(\agl{1}{3}\agl{1}{4}\sqr{3}{1}^2 + \agl{2}{3}\agl{2}{4}\sqr{3}{2}^2\right)\,,
    \\
    \mathcal A(\phi,\phi,q^-_{\alpha ip},\overline q^+_{\beta jr}) &= \frac{1}{2}\,\coef{8}{\phi^2 q^2}{pr}\left(\agl{1}{3}\agl{1}{4}\sqr{4}{1}^2 + \agl{2}{3}\agl{2}{4}\sqr{4}{2}^2\right)\delta_{\alpha\beta}\delta_{ij}\,,
    \\
    \mathcal A(\phi,\phi,u^+_{\alpha p},\overline u^-_{\beta r}) &= \frac{1}{2}\,\coef{8}{\phi^2 u^2}{pr}\left(\agl{1}{3}\agl{1}{4}\sqr{3}{1}^2 + \agl{2}{3}\agl{2}{4}\sqr{3}{2}^2\right)\delta_{\alpha\beta}\,,
    \\
    \mathcal A(\phi,\phi,d^+_{\alpha p},\overline d^-_{\beta r}) &= \frac{1}{2}\,\coef{8}{\phi^2 d^2}{pr}\left(\agl{1}{3}\agl{1}{4}\sqr{3}{1}^2 + \agl{2}{3}\agl{2}{4}\sqr{3}{2}^2\right)\delta_{\alpha\beta}\,,
    \\
    \mathcal A(\phi,\phi,\ell^-_{ip},\overline e^-_{r},\overline H_j) &= -\coef{8}{\phi^2\ell e H}{pr}\,\agl{1}{2}\agl{3}{4}\sqr{2}{1}\delta_{ij}\,,
    \\
    \mathcal A(\phi,\phi,q^-_{\alpha ip},\overline d^-_{\beta r},\overline H_j) &= -\coef{8}{\phi^2qd H}{pr}\,\agl{1}{2}\agl{3}{4}\sqr{2}{1}\delta_{\alpha\beta}\delta_{ij}\,,
    \\
    \mathcal A(\phi,\phi,q^-_{\alpha ip},\overline u^-_{\beta r}, H_j) &= -\coef{8}{\phi^2qu H}{pr}\,\agl{1}{2}\agl{3}{4}\sqr{2}{1}\delta_{\alpha\beta}\epsilon_{ij}\,.
\end{align}

\section{Positivity bounds in a UV extension}
\label{Appendix_B} 

To test the validity of the positivity bounds presented in Section~\ref{sec:positivity}, we match a UV theory to the ALP EFT.
A simple way to generate at tree level the operators subject to the positivity constraints consists in integrating out a massive spin-2 field $h_{\mu\nu}$ \textit{minimally} coupled to the stress-energy tensor $T_{\mu\nu}$ \cite{Fierz:1939ix,Hinterbichler:2011tt}:
\begin{equation}
    \mathcal L = \frac{1}{2}\partial_\lambda h_{\mu\nu}\partial^\lambda h^{\mu\nu}-\partial_\mu h_{\nu\lambda}\partial^\nu h^{\mu\lambda} +\partial_\mu h^{\mu\nu}\partial_\nu h-\frac{1}{2}\partial_\lambda h  \partial^\lambda h-\frac{M^2}{2}(h_{\mu\nu}h^{\mu\nu}-h^2)
    +\frac{c}{\Lambda}h^{\mu\nu}T_{\mu\nu}
\end{equation}
where $h=h^\mu_\mu$ is the trace of the field, $c$ is a real dimensionless coefficient, and $\Lambda$ is an energy scale much higher than the ALP EFT cutoff scale. Regarding the stress-energy tensor, we can consider
\begin{equation}
    T_{\mu\nu} = T_{\mu\nu}^{(\phi)} + T_{\mu\nu}^{(H)} + \sum_{\psi = \ell, e , q, u, d} T_{\mu\nu}^{(\psi)} + \sum_{X=B,W,G} T_{\mu\nu}^{(X)}
\end{equation}
with
\begin{align}
    T_{\mu\nu}^{(\phi)} &= \partial_\mu \phi \, \partial_\nu \phi - \frac{1}{2}g_{\mu\nu}\partial_\rho \phi \, \partial^\rho \phi\,,\label{eq:set_phi}\\
    T_{\mu\nu}^{(H)} &= D_\mu H^\dagger D_\nu H + D_\nu H^\dagger D_\mu H - g_{\mu\nu}D_\rho H^\dagger D^\rho H\,,\label{eq:set_higgs}\\
    T^{(\psi)}_{\mu\nu} &= \kappa^{(\psi)}_{pr}\left[\frac{i}{4}\left( \overline \psi_p \gamma_\mu \overset{\leftrightarrow}{D}_\nu \psi_r+\overline \psi_p \gamma_\nu \overset{\leftrightarrow}{D}_\mu \psi_r\right) -  g_{\mu\nu} \overline \psi_p i\slashed{D} \psi_r\right]\,,\\
    T^{(X)}_{\mu\nu} &= -X^{\mathscr A}_{\mu\rho} X_\nu^{\mathscr A \rho} + \frac{1}{4}g_{\mu\nu} X^{\mathscr A}_{\rho\sigma}X^{\mathscr A \rho\sigma}\,.
\end{align}
The matrices $\kappa^{(\psi)}$ must be Hermitian and positive definite ($\kappa^{(\psi)} \succ 0$) to ensure that unitarity is not violated.

Integrating out at tree level $h_{\mu\nu}$, we generate dimension-8 operators that belong both to the SMEFT and to the ALP EFT.
Concerning the latter, the associated Wilson coefficients take the following values:
\begin{gather}
    \coef{8}{\phi^4}{} =\frac{1}{3}\left( \frac{c}{\Lambda M}\right)^2\,,
    \\
    \coef{8}{\phi^2 H^2}{(1)} =-\frac{2}{3}\left( \frac{c}{\Lambda M}\right)^2\,, \qquad \coef{8}{\phi^2 H^2}{(2)} =2\left( \frac{c}{\Lambda M}\right)^2\,,
    \\
    \coef{8}{\phi^2 X^2}{(1)} =-\left( \frac{c}{\Lambda M}\right)^2\,, \quad 
    \coef{8}{\phi^2 X^2}{(2)} =\frac{1}{4}\left( \frac{c}{\Lambda M}\right)^2\,,\quad 
    \coef{8}{\phi^2 X^2}{(3)} =0 \,,
    \quad X=B,W,G\,,
    \\
    \coef{8}{\phi^2 \psi^2}{pr} =\kappa^{(\psi)}_{pr}\left( \frac{c}{\Lambda M}\right)^2\,,
    \qquad \psi = \ell,e,q,u,d\,.
\end{gather}
All of them are consistent with the positivity constraints of Section~\ref{sec:positivity}.

The same conclusion holds if $h_{\mu\nu}$ is \textit{conformally} coupled, which consists in substituting the expressions for $T_{\mu\nu}^{(\phi)}$ and $T_{\mu\nu}^{(H)}$ in Eqs.~\eqref{eq:set_phi} and \eqref{eq:set_higgs} with the improved traceless combinations~\cite{Callan:1970ze}
\begin{align}
    T_{\mu\nu}^{(\phi)} &= \partial_\mu \phi \, \partial_\nu \phi - \frac{1}{2}g_{\mu\nu}\partial_\rho \phi \, \partial^\rho \phi - \frac{1}{6}(\partial_\mu \partial_\nu - g_{\mu\nu}\partial_\rho \partial^\rho)\phi^2\,,\\
    T_{\mu\nu}^{(H)} &= D_\mu H^\dagger D_\nu H + D_\nu H^\dagger D_\mu H - g_{\mu\nu}D_\rho H^\dagger D^\rho H \nonumber \\ &\quad - \frac{1}{6}(D_\mu D_\nu+D_\nu D_\mu -2 g_{\mu\nu}D_\rho D^\rho)(H^\dagger H)\,,
\end{align}
respectively.
With respect to the previous case, none of the Wilson coefficients change value, and the effective Lagrangian remains consistent with the positivity bounds.

\bibliographystyle{JHEP}
\bibliography{ALP_bibliography}

\end{document}